\documentclass[iop,revtex4]{emulateapj}

\slugcomment{DRAFT: \today}
\shorttitle{Near-IR Spectral Imaging of the Orion Fingers}
\shortauthors{Youngblood et al.}
\received{January 6, 2016}
\revised{March 14, 2016}
\accepted{April 13, 2016}

\usepackage{subfigure}
\usepackage{color}
\usepackage{mathtools}
\usepackage[hidelinks]{hyperref}

\hypersetup{ colorlinks, citecolor=black, filecolor=black, linkcolor=blue, urlcolor=blue }

\newcommand{\HH}{H$_{2}$}
\newcommand{\Msun}{M$_{\odot}$}

\newcommand{\mum}{\ensuremath{\mu \mathrm{m}}}
\newcommand{\FeII}{[\ion{Fe}{2}]}
\newcommand{\FeIII}{[\ion{Fe}{3}]}
\newcommand{\OI}{[O\,\textsc{i}]}
\newcommand{\HI}{H\,\textsc{i}}
\newcommand{\HII}{\ion{H}{2}}
\newcommand{\HeI}{He\,\textsc{i}}
\newcommand{\kms}{km s$^{-1}$}
\newcommand{\BrG}{Br$\gamma$}
\newcommand{\PII}{[\ion{P}{2}]}

\begin{document}

\title{The Orion fingers: Near-IR spectral imaging of an explosive outflow
        }

\author{ Allison Youngblood$^1$, Adam Ginsburg$^2$, John Bally$^1$ }
\altaffiltext{1}{Department of Astrophysical and Planetary Sciences, University of Colorado, UCB 389, Boulder, CO 80309, USA}
\altaffiltext{2}{ESO Headquarters, Karl-Schwarzschild-Str. 2, 85748 Garching bei M{\"u}nchen, Germany}

\begin{abstract}

We present near-IR (1.1--2.4 \mum) position-position-velocity cubes of the 500-year-old Orion BN/KL explosive outflow with spatial resolution 1\arcsec~and spectral resolution 86 \kms.  We construct integrated intensity maps free of continuum sources of 15 \HH~and \FeII~lines while preserving kinematic information of individual outflow features.  Included in the detected \HH~lines are the 1-0 S(1) and 1-0 Q(3) transitions, allowing extinction measurements across the outflow.  Additionally, we present dereddened flux ratios for over two dozen outflow features to allow for the characterization of the true excitation conditions of the BN/KL outflow.  All ratios show the dominance of shock excitation of the \HH~emission, although some features exhibit signs of fluorescent excitation from stellar radiation or J-type shocks.  We also detect tracers of the PDR/ionization front north of the Trapezium stars in \OI~and \FeII~and analyze other observed outflows not associated with the BN/KL outflow.

\end{abstract}
\section{Introduction} \label{sec:Introduction}

The Orion BN/KL outflow, located at 414 $\pm$~7 pc \citep{Menten2007}, just behind the Orion Nebula, is an explosive, wide-angle outflow emerging from the OMC1 cloud core known for its bright, shock-excited \HH~and \FeII~emission and several reflection nebulae.  8 $\pm$~4 \Msun~of material is entrained in the inner, slow-moving (20 \kms) part of the outflow \citep{Snell1984} with less than 1 \Msun~in the high-velocity (150--300 \kms) fingers and bullets.  Approximately 120 different jet-like structures are seen all with similar dynamic ages, indicating that they originate from a single event \citep{Bally2015}. In contrast to a grouping of traditional young stellar object (YSO) outflows, the velocity structure of the BN/KL outflow is consistent with a Hubble type flow indicating an explosive origin \citep{Zapata2009,Bally2015}.  

Radio proper motion measurements show that the BN object, source I, and possibly source n were within 500 AU from each other at (05:35:14.360, -05:22:28.70) approximately 500 years ago \citep{Goddi2011, Gomez2008, Gomez2005, Rodriguez2005}, and the near-IR bowshocks trace back to a common point within a few arcseconds of the ejection center \citep{Bally2011, Bally2015}.  Possible launch mechanisms include the dynamical decay of an unstable multiple system of young stars \citep{Bally2015, Zapata2009}, a merger of massive stars \citep{Bally2005}, or a period of intense accretion onto source I caused by the close passage of a runaway massive star \citep{Tan2004, Chatterjee2012}.  

All three proposed origins of the BN/KL outflow indicate this phenomenon might be common to massive star forming regions (SFRs), and other possible examples include include DR21 \citep{Zapata2013}, G34.26+0.15 \citep{Cyganowski2008}, NGC 7129 \citep{Eisloffel2000,Gutermuth2004}, IRAS 05506+2414 \citep{Sahai2008}, and W49 Source G \citep{Smith2009}.  The BN/KL outflow is the closest and thus best for detailed observations diagnosing the conditions and properties of this type of outflow.

Observations of multiple transitions of \HH~can be used to derive excitation conditions of shocks on a pixel-by-pixel basis (e.g., \citealt{Colgan2007}).  \HH~emission is excited collisionally with other \HH~molecules, \HI, and electrons in hot 2000 K post-shock gas.  These collisions populate only the lower vibrational levels in the ground electronic state, and $\Delta$$v$ = 1 transitions resulting from quadrupolar radiative decay emit primarily in the K band \citep{Wolfire1991}.  However, UV photons from the Lyman and Werner bands can excite \HH~molecules to the first excited electronic state where the molecule can then dissociate (10\% of the time) or decay into bound ro-vibrational levels of the electronic ground state resulting in a cascade of transitions down to the ground level \citep{Shull1978,Black1976,Black1987}.  These two excitation mechanisms create distinct signatures in the flux ratios of the resulting K band transitions.  For example, a $\sim$10:1 flux ratio in the 2.12 \mum~1-0 S(1) to 2.24 \mum~2-1 S(1) lines is expected for shocks (12 assuming LTE; \citealt{Marconi1998}) because only the lower vibrational levels are populated.  UV-excited \HH~molecules can populate both high- and low-$v$ states in the electronic ground state, decreasing the flux ratio in the 2.12 \mum~1-0 S(1) to the 2.24 \mum~2-1 S(1) line to $\sim$~2 \citep{Black1976,Black1987,Wolfire1991}.

We present medium-resolution, near-IR spectroscopy used to construct position-position-velocity (PPV) maps of the available \HH~and \FeII~lines in the 1.1--2.4 \mum~range at $\sim$~1\arcsec~spatial resolution and 86 \kms~spectral resolution. Our constructed integrated intensity maps for the observed spectral lines exclude significant contamination by stars and reflection nebulosity and preserve kinematic information. We characterize the excitation conditions of the shocked gas and determine the kinematics of the wide-angle BN/KL outflow.  In the optical, similar work has been presented by \cite{Doi2004}, \cite{Garcia-Diaz2007}, and \cite{Garcia-Diaz2008} at higher spectral resolution and over a larger area of the Orion Nebula, but the optical only traces the foreground \HII~region.

Section~\ref{sec:ObservationsReductions} presents the observations and data reduction. Section~\ref{sec:Results} describes the results, including the \HH~and \FeII~emission, kinematics, visual extinction, \HH~excitation, and YSO jets present in the field. Section~\ref{sec:Summary} summarizes the results.

\section{Observations \& Reductions} \label{sec:ObservationsReductions}

Spectra were obtained on 5, 6, 24, and 25 November 2012, and 24 February 2013 using the cross-dispersed near-IR TripleSpec spectrograph on the Apache Point Observatory (APO) 3.5-meter telescope with the 1.1\arcsec~$\times$  43\arcsec~slit.  TripleSpec has a resolution $\lambda$/$\Delta \lambda$ = 3500 with the 1.1\arcsec~ slit and a dispersion of 0.39\arcsec~pix$^{-1}$.  Using a mapping script, we stepped across the 2.7\arcmin~$\times$~3.3\arcmin~Orion BN/KL outflow in 0.5\arcsec~increments to ensure full spatial coverage. The outflow was scanned from north to south and east to west typically in 80-step scans. Spectra of the A0V standard star HD 37547 were obtained each night for flux calibration in pairs at two locations along the slit (ABBA nods).  All integration times were 30 seconds.
\begin{figure*}
     \begin{center}

        \subfigure{
           
            \includegraphics[width=\textwidth]{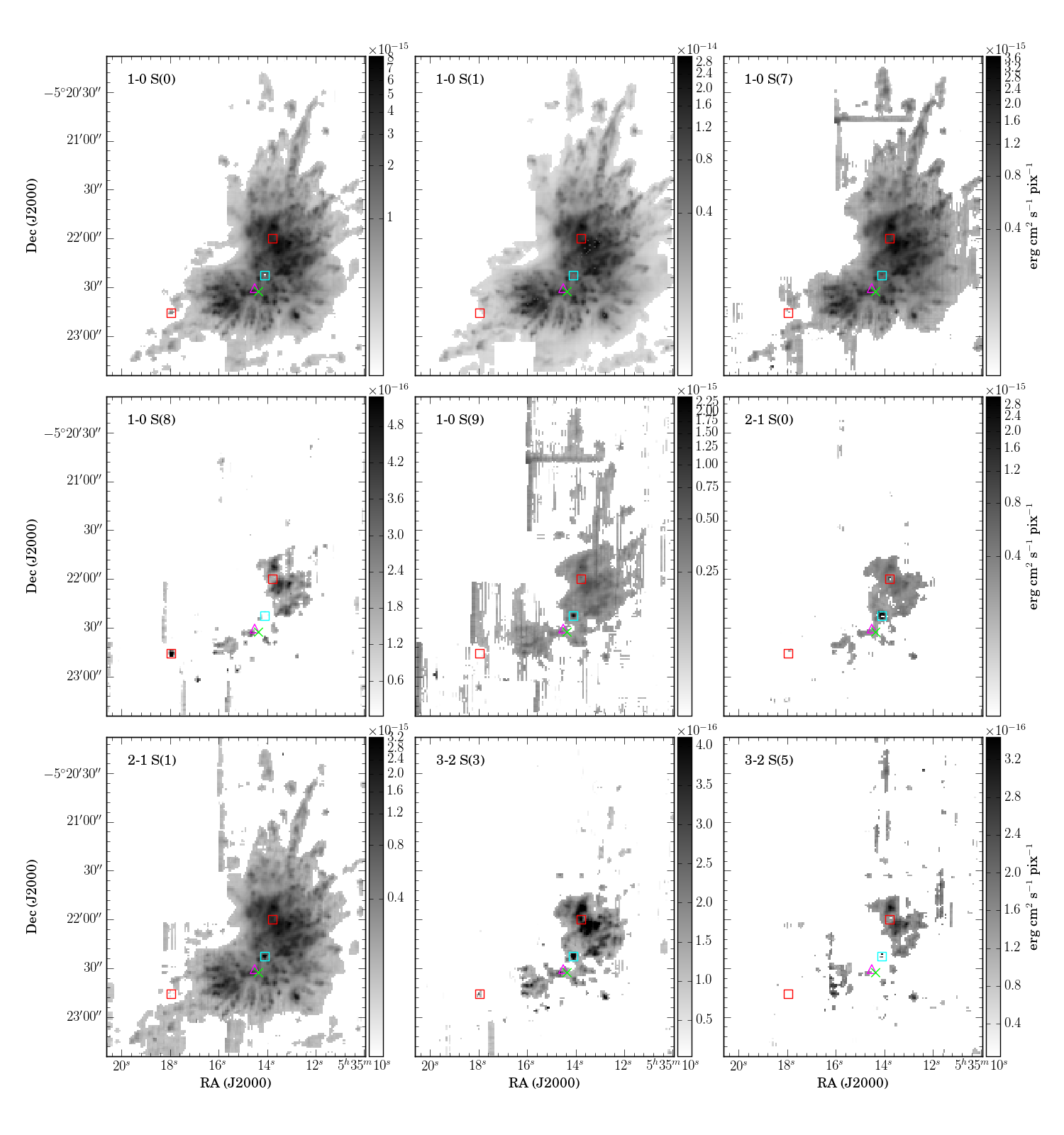}
        }
       
    \end{center}
    \caption{
        \HH~S branch integrated intensity maps in erg cm$^{-1}$ s$^{-1}$ pix$^{-1}$. Pixels with S/N \textless~10 are masked. The cyan square, magenta triangle, and green X mark the locations of the BN object, source I, and source n, respectively. The red squares mark the locations of V2248 Ori (northwest) and MT Ori (southeast).
     }
   \label{fig:h2_sbranch_series}
\end{figure*}

\begin{figure*}
     \begin{center}

        \subfigure{
          \includegraphics[width=\textwidth]{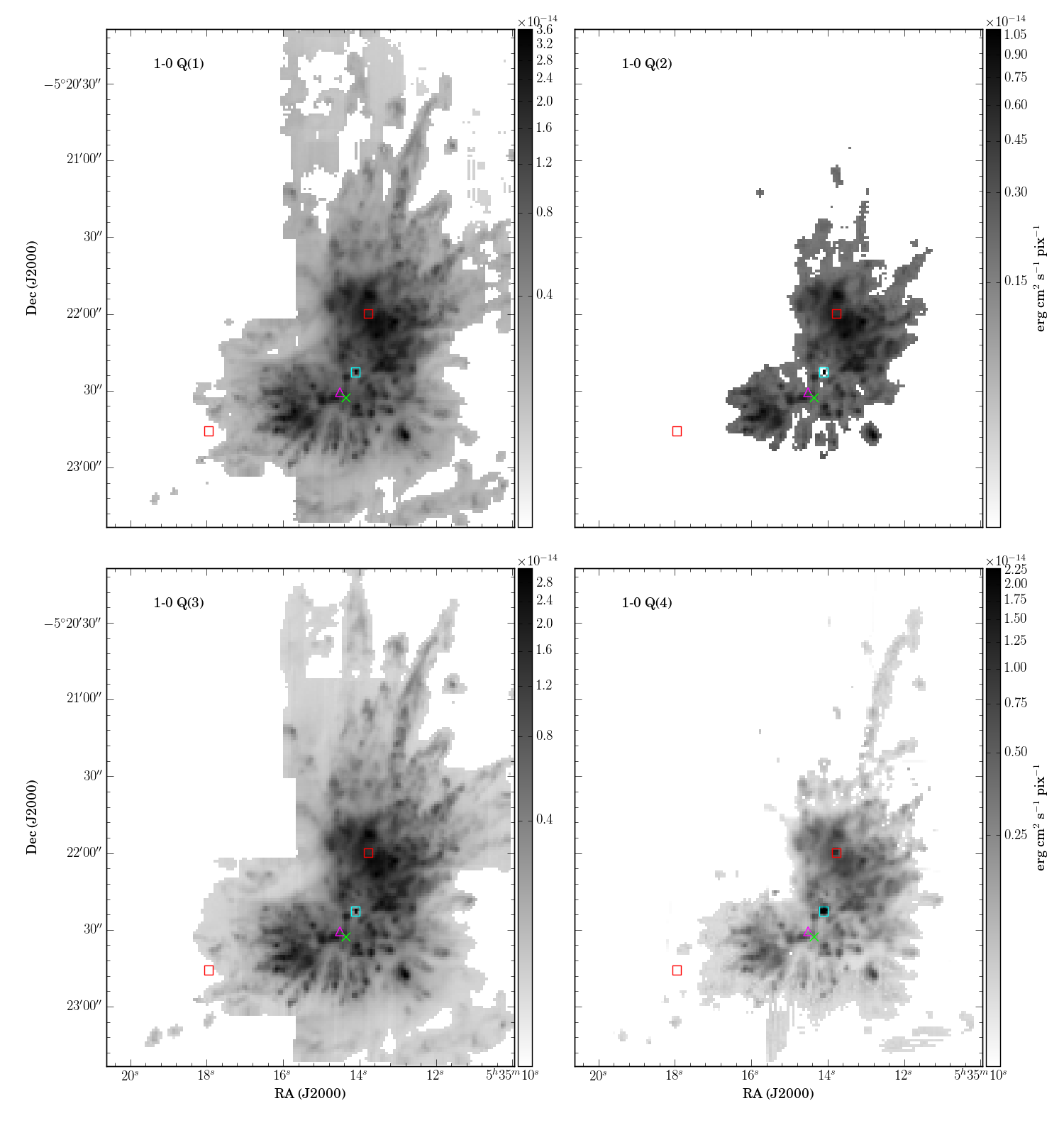}
        }

    \end{center}
    \caption{
        \HH~Q branch integrated intensity maps in erg cm$^{-1}$ s$^{-1}$ pix$^{-1}$. Pixels with S/N \textless~10 are masked. The cyan square, magenta triangle, and green X mark the locations of the BN object, source I, and source n, respectively. The red squares mark the locations of V2248 Ori (northwest) and MT Ori (southeast).}
    
   \label{fig:h2_qbranch_series}
\end{figure*}

Standard IRAF\footnote{IRAF is distributed by the National Optical Astronomy Observatories, which are operated by the Association of Universities for Research in Astronomy, Inc., under cooperative agreement with the National Science Foundation.} tasks were used for the initial reduction. First, cosmic rays were removed from all the raw spectra.  Then, the orders (J, H, and K) were cropped and rectified in the spectral direction using the A and B nod positions of the standard star and in the spatial direction using the airglow lines.  The airglow lines were also used for wavelength identification \citep{Rousselot2000,Lord1992,Cushing2004}.  All wavelengths are reported in vacuum.  Due to flexure in the instrument over the course of the night, the spectra were placed in groups where the airglow lines matched in wavelength space, and those spectra were cropped, rectified, and wavelength calibrated as a single group.  The J, H, and K orders were then stitched together, spatially flat fielded, and the top and bottom three pixel rows were cropped due to vignetting.  Because the fit to the standard star trace is discrete, a velocity gradient in the direction of the slit scan was introduced in the spectra.  This artifact of the reduction process will be further discussed in Section~\ref{sec:HHfingers}.

The background in each spectrum was subtracted by masking all emission lines and continuua, and fitting a 2nd order polynomial to each row in the spectral direction and then subtracting.  Then, airglow lines that do not lie within a few spectral-axis pixels of an emission line of interest were median subtracted, and all other airglow lines were not subtracted so as not to affect the target emission lines.  Airglow lines close to or blended with a source emission line are removed later in the integrated intensity map making process.

A telluric correction spectrum was created using APO TripleSpecTool \citep{Vacca2003} and the A0V standard star HD 37547.  The correction spectrum was applied, and each spectrum was shifted into the kinematic Local Standard of Rest.  Because TripleSpec has moderate resolution (86 \kms), only a single velocity correction was applied per night of data.  The individual slit scans were combined into position-position-velocity (PPV) data cubes and coordinates were applied via IRAF's ccmap, cctran, and ccsetwcs tasks.  By comparing the positions of the stars in the PPV cubes to a 2010 APO NICFPS \HH~1-0 S(1) image, we estimated that the World Coordinate System (WCS) solutions of each cube are accurate to within 1\arcsec. Based on the root-mean-square of the residuals from the wavelength solutions, we estimate that the wavelength solutions are accurate to within 1 \AA~(25 \kms) in J band, 0.5 \AA~(10 \kms) in H band, and 1.2 \AA~(17 \kms) in K band. The PPV cubes were then averaged together into one cube\footnote{The J, H, and K band PPV cubes are available for download at http://dx.doi.org/10.7910/DVN/YUNZ1F. The reduction scripts are available at https://github.com/allisony/TspecCubes.}.  The final dispersion is 2.88 \AA~pix$^{-1}$ and the image scale is 1\arcsec~pix$^{-1}$.  

Integrated intensity maps of the emission lines were made using the \texttt{pyspeckit} Python module \citep{Ginsburg2011} which fits single Gaussians\footnote{By default \texttt{pyspeckit} uses the Levenberg-Marquardt algorithm via MPFIT \citep{Markwardt2009}.} (background level, amplitude, $\sigma$, and velocity centroid as free parameters) to individual emission lines.  The emission line flux was calculated analytically as the integral of a Gaussian ($\sqrt{2\pi}$ A $\times$ $\sigma$).  If the line was blended with an airglow line or had one or more within a few spectral-axis pixels, extra Gaussians were fit simultaneously and the original emission line flux was recovered.  There are a few transitions of interest that completely coincide with airglow, and we were not able to recover the original emission flux.  The free parameters of the Gaussian fit allow us to reconstruct mostly pure emission line maps with no stars or other continuum sources while retaining kinematic information from the velocity centroid and sigma. The exception is three bright stars (BN object, V2248 Ori, and MT Ori) erroneously contribute emission in some of our integrated intensity maps. They have been marked in many of the figures for easy identification of stellar contamination.

Fits were made only for lines above a signal-to-noise (S/N) threshold of 10. Linewidths ($\sigma$) were constrained between 1 \AA~and 5 \AA.  To replace missing pixels where no fits occurred, we used 2D interpolation over the integrated intensity maps. Uncertainties in the fluxes are typically no more than 30\%.

\begin{figure*}
     \begin{center}

        \subfigure{
         
           \includegraphics[width=\textwidth]{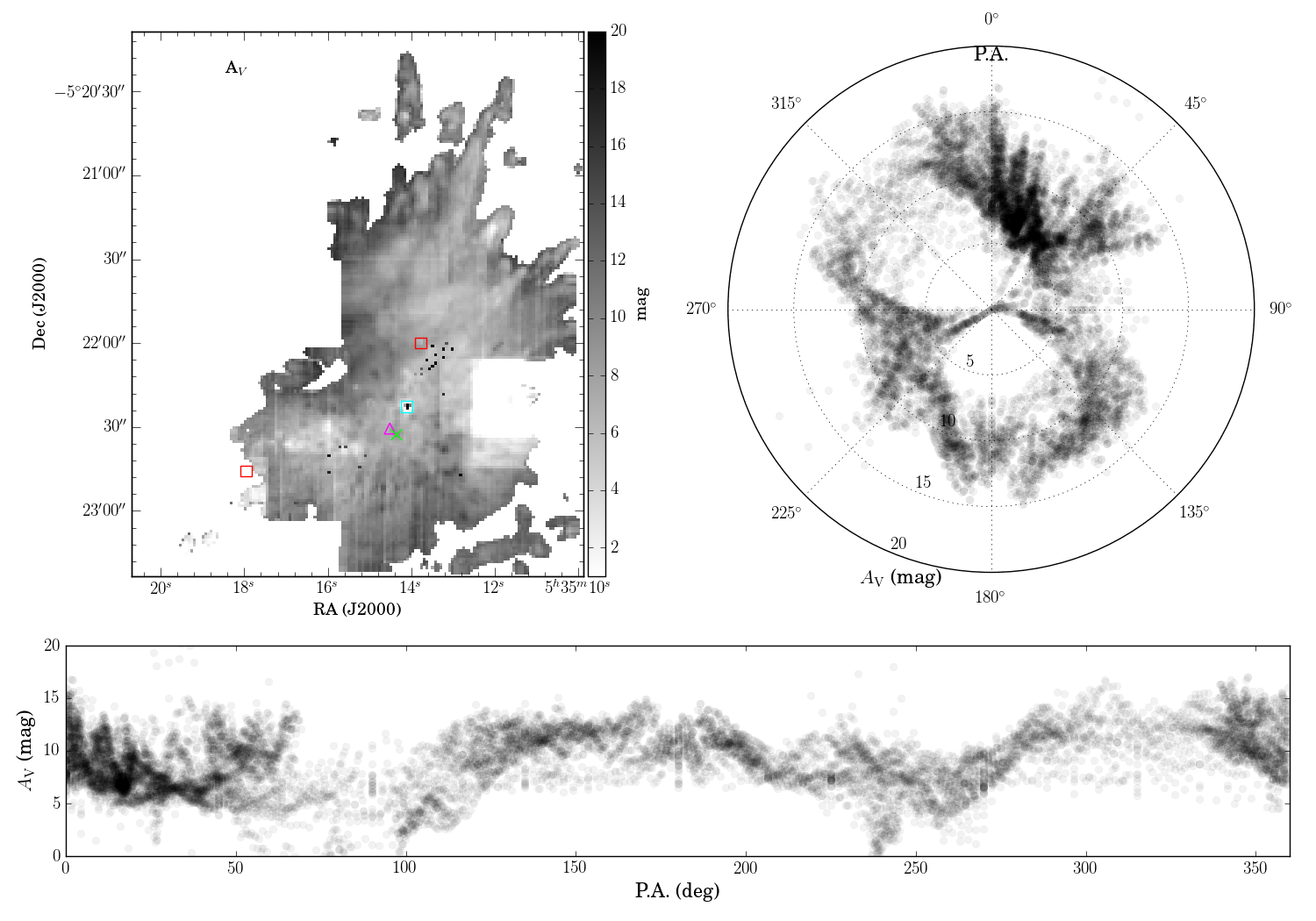}
         }

    \end{center}
    \caption{
        ($Top$ $Left$) Visual extinction ($A_{\rm V}$) map made from \HH~1-0 Q(3) and 1-0 S(1) integrated intensity maps. The cyan square, magenta triangle, and green X mark the locations of the BN object, source I, and source n, respectively. The red squares mark the locations of V2248 Ori (northwest) and MT Ori (southeast). ($Top$ $Right$) $A_{\rm V}$ (radial axis) as a function of position angle (azimuthal direction). The center is defined as the center of the outflow (05:35:14.360, 05:22:28.70). PA = 0\,$^{\circ}$ is North and PA = 270\,$^{\circ}$ is East. ($Bottom$) Similarly to the top right panel, $A_{\rm V}$ as a function of position angle.
     }
   \label{fig:AV_polar}
\end{figure*}

\begin{figure*}
     \begin{center}

        \subfigure{
         
           \includegraphics[width=\textwidth]{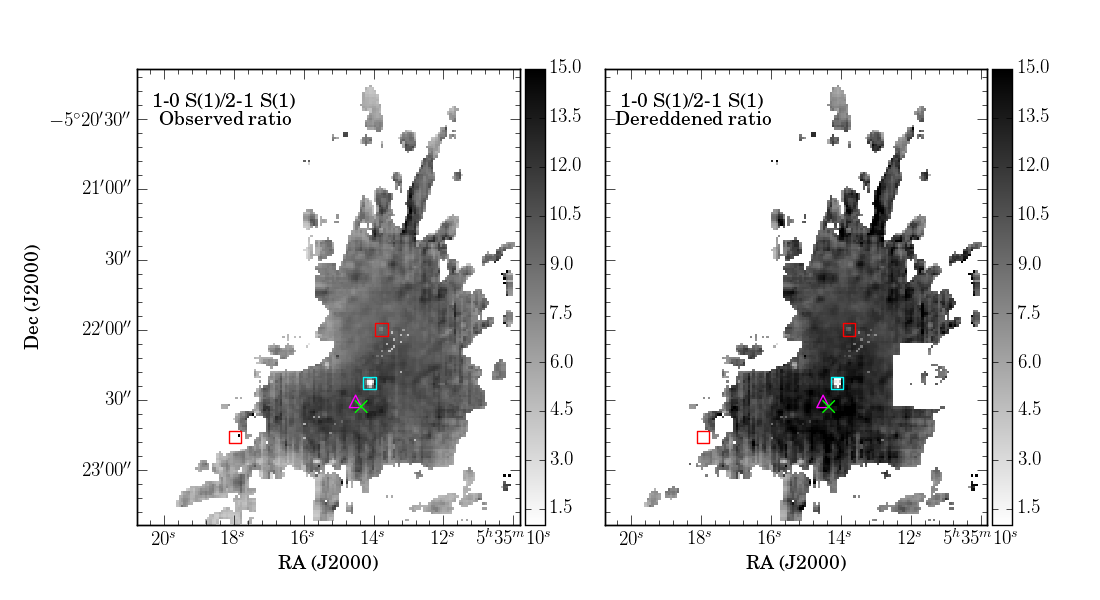}
         }

    \end{center}
    \caption{
        ($Left$) Flux ratio map of the \HH~1-0 S(1) 2.12 \mum~and 2-1 S(1) 2.25 \mum~lines.  ($Right$) The same as the left panel, but corrected for extinction using the $A_{\rm V}$ map in Figure~\ref{fig:AV_polar}. In all three panels, the cyan square, magenta triangle, and green X mark the locations of the BN object, source I, and source n, respectively. The red squares mark the locations of V2248 Ori (northwest) and MT Ori (southeast).
     }
   \label{fig:H2ratio}
\end{figure*}

\begin{figure*}
     \begin{center}

        \subfigure{
          
           \includegraphics[width=\textwidth]{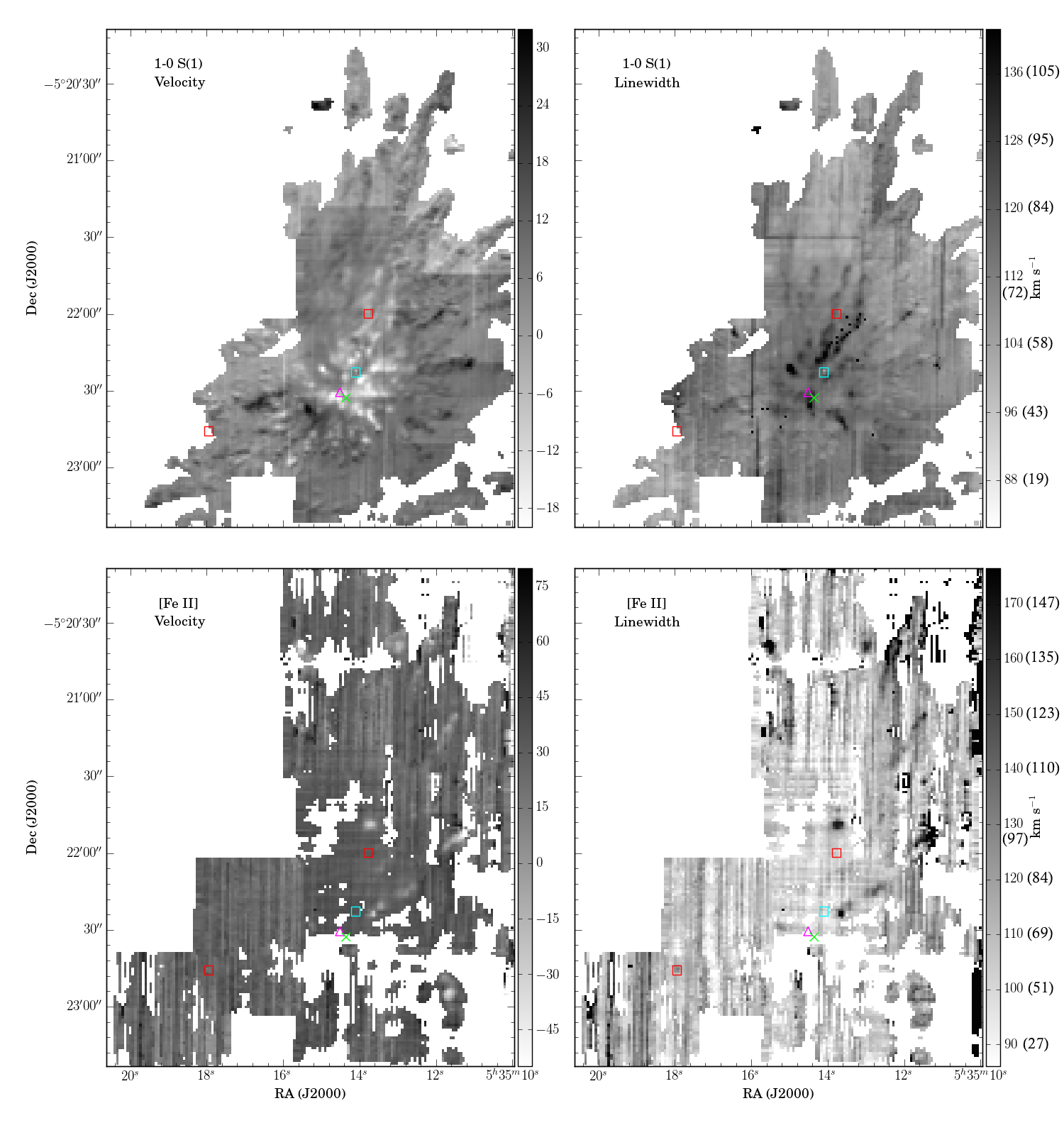}
         }

    \end{center}
    \caption{
      Velocity and linewidth (FWHM) maps derived from Gaussian fits for \HH~1-0 S(1) at 2.12 \mum~(top row) and \FeII~1.26 \mum~(bottom row). The values in parentheses on the two linewidth colorbars show the deconvolved linewidths. The cyan square, magenta triangle, and green X mark the locations of the BN object, source I, and source n, respectively. The red squares mark the locations of V2248 Ori (northwest) and MT Ori (southeast).
     }
   \label{fig:kinematics}
\end{figure*}

Absolute flux calibration with long-slit spectroscopy is not possible because an unknown fraction of the standard star's light falls outside the slit.  To account for this missing flux, we made \HH~1-0 S(1) 2.12 \mum~integrated intensity maps from each of our individual slit scans and compared them to a 2010 APO NICFPS \HH~1-0 S(1) flux-calibrated image.  We used the ratio of the two images to increase the flux appropriately in each of the data cubes before averaging them together.  The typical correction factor ranged from 1.4--3.4 with an average of 2.

\section{Results} \label{sec:Results}

The PPV cubes presented here provide the first wide-field near-IR view of the Orion BN/KL outflow in dozens of spectral lines.  We utilize integrated intensity maps of \HH~and \FeII~to diagnose the kinematics and excitation conditions of the region.

\subsection{\HH~Fingers} \label{sec:HHfingers}

The famous \HH~fingers of the BN/KL outflow are visible and unaffected by significant telluric absorption in 22 ro-vibrational \HH~lines in our TripleSpec spectra. However, approximately half these transitions are coincident with telluric emission (airglow), making them only useful for determining the morphology of the emission in that line.  Figures~\ref{fig:h2_sbranch_series} and~\ref{fig:h2_qbranch_series} show \HH~$v$ = 1$\rightarrow$0 S branch ($\Delta J$ = +2) and $v$ = 1$\rightarrow$0 Q branch ($\Delta J$ = 0) integrated intensity maps free of continuum sources.  The positions of the BN object, source I, and source n are denoted in each figure by a blue square, magenta triangle, and green X, respectively. The BN object, V2248 Ori (05:35:13.8,-5:21:59.6) and MT Ori (05:35:17.9,-5:22:45.4) sometimes contaminate the pure emission line maps (see Section~\ref{sec:ObservationsReductions}).  The quality of the maps depends jointly on the S/N of the line and airglow contamination. Many of the lines suffer from at least partial contamination (within a few pixels of the line core), and the higher S/N lines allow for easier simultaneous fitting of the source line and airglow contaminant.

The greatest \HH~intensity occurs in the central 85\arcsec~$\times$~75\arcsec~region -- Peak 1 and Peak 2 \citep{Beckwith1978} -- where the proper motion of the outflow is the slowest \citep{Bally2015}. The fingers with the greatest proper motions (150--300 \kms) extend 1.5\arcmin--2.5\arcmin~away from the center of the outflow. Figure~\ref{fig:kinematics} shows the intensity-weighted radial velocities ($v_{\rm LSR}$) and linewidths of the \HH~1-0 S(1) line derived from the \texttt{pyspeckit} fits. The central region of the flow is generally blueshifted and exhibits broader linewidths, while the outer fingers are typically redshifted (see Section~\ref{kinematics}). The fast-moving fingers are primarily in the plane of the sky because the proper motions are more than twice the line-of-sight motion, while the the slow-moving fingers are not.

\subsubsection{Visual extinction \label{sec:AV}}
To recover the intrinsic emission from the observed \HH~line emission and derive the true excitation pixel-by-pixel, we constructed a visual extinction ($A_{\rm V}$) map from the \HH~1-0 Q(3) and 1-0 S(1) lines.  These two lines originate from the same upper state ($v$ = 1, $J$ = 3) and have an intrinsic line ratio of 0.74 \citep{Geballe1982}.  We assumed a typical near-IR extinction law from \cite{Cardelli1989} to derive $A_{\rm V}$ (Figure~\ref{fig:AV_polar}).   \HH~1-0 Q(3) is partially blended with at least one airglow line, and as a result, the individual slit scans with severe airglow contamination were excluded from the $A_{\rm V}$ map.  $A_{\rm V}$ is as low as 3.5 mag in the southeastern part of the outflow, and as high as 35 mag in the northern fingers.  In the central regions, the extinction is typically 20 mag.  Regions with low S/N typically over-predict $A_{\rm V}$ because airglow contamination dominates the 1-0 Q(3) line there.  Assuming a typical gas-to-dust ratio of 100, the \HI~+ \HH~column density in the central region is $\sim$4$\times$10$^{22}$ cm$^{-2}$, consistent with the \HH~emission originating inside the OMC1 cloud.

Figure~\ref{fig:AV_polar} also shows the visual extinction as a function of position angle (PA). The center is the origin of the outflow (05:35:14.360, -05:22:28.70) and PA = 0\,$^{\circ}$ and 270\,$^{\circ}$ are north and east, respectively. Each point represents a single pixel in the $A_{\rm V}$ map in the left panel of Figure~\ref{fig:AV_polar}. Along the outflow axis (NW -- SE), large and small extinction values are present, while the orthogonal axis contains no pixels with $A_{\rm V}$ $\lesssim$~6. This NE -- SE axis corresponds to the PA of the Orion A integral-shaped filament \citep{Salji2015}, so we do not expect low extinction values.

\subsubsection{Shock excitation \label{shockexcitation}}
Figure~\ref{fig:H2ratio} shows the ratio between the 1-0 S(1) and 2-1 S(1) lines, a common shock diagnostic, before extinction correction and after. In the extinction-corrected ratio map, the pixels with ratios \textless~10 indicate some UV excitation, although it is not dominant.  \cite{Goicoechea2015} and \cite{Chen2014} confirm the far-UV irradiation of the shocked material in the BN/KL outflow but cannot distinguish between an external source like the Trapezium or internal illumination by J-type shocks.  Table~\ref{table:H2_fluxes} presents the observed flux ratios of various knots highlighted in Figure~\ref{fig:regions}, and Table~\ref{table:H2_fluxes_dered} presents the de-reddened flux ratios for the same knots.  For the selected knots, the 1-0 S(1)/2-1 S(1) ratio varies from 6--12.5, indicating shocks are the dominant excitation mechanism of the \HH~molecules.

\subsubsection{Kinematics \label{kinematics}}
The radial velocity map shows the central region of the BN/KL outflow is generally blueshifted and the outer fingers are generally redshifted (Figure~\ref{fig:kinematics}). However, CO observations of the central region of the outflow show roughly the same number of redshifted and blueshifted streamers \citep{Zapata2009}, indicating the outflow may be isotropic. The receding \HH~fingers in the central region are probably hidden by dense gas ($A_{\rm V}$ $\approx$~20 mag) from the OMC1 hot core. The \HH~linewidths in the central region are large (FWHM\,$\approx$\,115--140 \kms; deconvolved FWHM\,$\approx$\,76--110 \kms), and the expected radial velocity differences of the blueshifted and redshifted flows are of order TripleSpec's 86 \kms~spectral resolution. It is possible that blueshifted and redshifted \HH~fingers coincident on the plane-of-the-sky are simply not spectrally resolved. This is especially likely if the redshifted fingers are much fainter than the blueshifted fingers, because our single Gaussian fits' velocity centroids will be skewed to the brighter velocity components. The redshifted fingers, as mentioned above, are likely to suffer from more extinction than the blueshifted fingers. Also, the CO observations \citep{Zapata2009} generally show radial velocities $v_{\rm LSR}$ \textless~-30 \kms~and $v_{\rm LSR}$ \textgreater~+30 \kms, but our \HH~radial velocities are -30 \kms~\textless~$v_{\rm LSR}$ \textless~+30 \kms. This indicates the near-IR \HH~shocks may not be tracing the same gas as seen in CO.

\begin{figure*}
     \begin{center}

        \subfigure{
         
           \includegraphics[width=\textwidth]{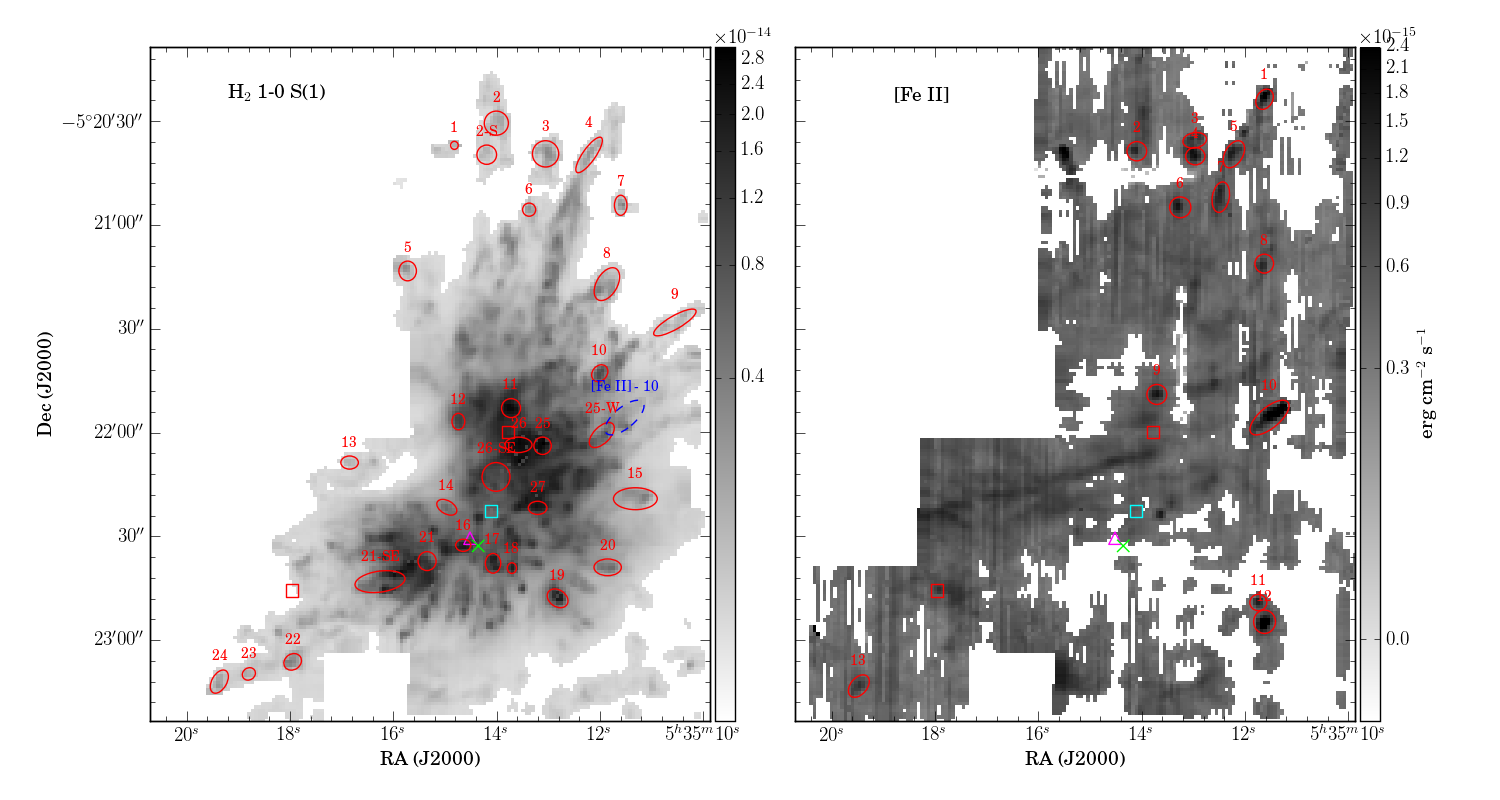}
         }
         
            \end{center}
    \caption{
       ($Left$) \HH~1-0 S(1) integrated intensity map with the red elliptical regions showing the features marked in the Tables~\ref{table:H2_coords}--\ref{table:H2_fluxes_dered}.  ($Right$) \FeII~1.26 \mum~integrated intensity map with the red elliptical regions showing the features listed in Table~\ref{table:FeII_fluxes}. The cyan square, magenta triangle, and green X mark the locations of the BN object, source I, and source n, respectively. The red squares mark the locations of V2248 Ori (northwest) and MT Ori (southeast).
     }
   \label{fig:regions}
\end{figure*}

The radial velocity and linewidth (FWHM) maps (Figure~\ref{fig:kinematics}; calculated from the single Gaussian fits' $\lambda_{0}$ and $\sigma$~parameters) combined with the $A_{\rm V}$ map (Figure~\ref{fig:AV_polar}) reveal kinematically distinct outflow features that would not be apparent from the integrated intensity maps alone.  Figure~\ref{fig:regions} and Table~\ref{table:H2_coords} highlight specific outflows with unique properties. Feature \#14 (see Figure~\ref{fig:regions} for numbered features) with the largest linewidth of 150 \kms~(deconvolved FWHM = 122 \kms) is a blueshifted $v_{\rm LSR}$ = -8 \kms~knot located at (05:35:14.961, -5:22:21.87).  With a \HH~1-0 S(1) total flux of 7.25$\times$10$^{-14}$ erg s$^{-1}$ cm$^{-2}$, it is among the dimmer knots.  At (05:35:15.348, -5:22:37.48) lies the significantly redshifted ($v_{\rm LSR}$ = +30 \kms) knot \#21 that also stands out kinematically because it is slightly broader than its surroundings.  Like the previous knot, it is almost indistinguishable from its surroundings in \HH~intensity.  

Knot \#21's $A_{\rm V}$ $\sim$~10 mag, is distinctly smaller than the surrounding area's $A_{\rm V}$ $\sim$~20--25 mag, suggesting that this knot is either on the far side of the flow and is only visible through a cavity or is an unrelated foreground flow. The area to the east of knot \#21 also shows $A_{\rm V}$ lower than its surroundings, and the shape of the low $A_{\rm V}$ values corresponds to an \HH~finger \#21-E at (5:35:16.681, -5:22:36.01) seen at $v_{\rm LSR}$ = +20 \kms~redshift. $A_{\rm V}$ is also relatively small for the \HH~fingers in the southeast and the west.  Knots of anomalous velocity with respect to the surrounding \HH~finger typically correspond to regions of low extinction, such as the knot \#2-S on the northern-most finger at (5:35:14.164, -5:20:39.24). \#2-S also corresponds to an \FeII~bullet. Near the center of the outflow lies an x-shaped feature \#26-SE (5:35:14,  -5:22:13.72) of low extinction that corresponds to a redshifted feature.

As mentioned in Section~\ref{sec:ObservationsReductions}, the discrete fit to the standard star trace introduced a velocity gradient into the PPV cube in the direction of the slit scans.  This effect can be seen in Figure~\ref{fig:kinematics} where many of the \HH~fingers show blueshifted emission to the north, and redshifted emission to the south.  Other fingers where this effect is not seen have data coverage from two slit scans in orthogonal directions.  No region of the cube has only east-west slit scan coverage, hence when the velocity gradient is seen, it is only in the north-south direction.

\subsection{[Fe\,\textsc{II}]~Fingertips} \label{sec:FeIIfingertips}

\begin{figure*}
     \begin{center}

 \subfigure{
           \includegraphics[width=\textwidth]{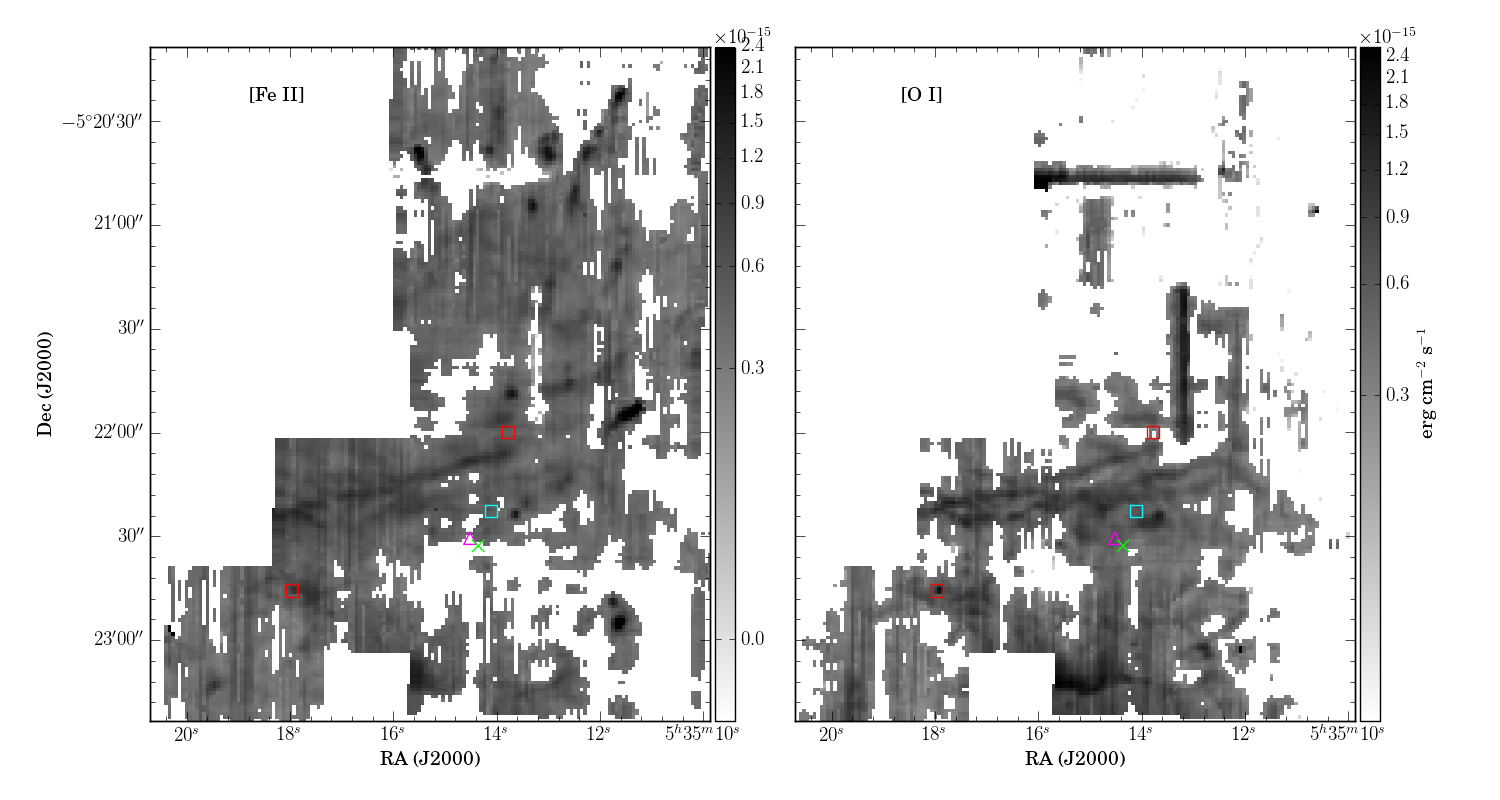}
         }

    \end{center}
    \caption{
        ($Left$) \FeII~1.26 \mum~integrated intensity map.  ($Right$) \OI~1.32 \mum~integrated intensity map. The cyan square, magenta triangle, and green X mark the locations of the BN object, source I, and source n, respectively. The red squares mark the locations of V2248 Ori (northwest) and MT Ori (southeast).
     }
   \label{fig:feii_series_pyspeckit}
\end{figure*}

We detect bright \FeII~emission lines at 1.26 \mum~and 1.64 \mum~and their related transitions originating from the $^{4}D$ term (second excited state) and terminating in the $^{6}D$ (ground state) and $^{4}F$ (first excited state) terms, respectively. The \FeII~emission is seen only around the shock heads (the fingertips) as the $^{4}$D term requires 12000 K collisional excitation \citep{Smith2006}.  Also, \FeII~emission is likely enhanced when grain sputtering via shocks or shock-heating creates a partially ionized zone, thus increasing the gas phase abundance of iron \citep{Greenhouse1991,Mouri2000}. The fingertip \FeII~emission appears primarily in the outer, faster-moving fingers in the plane of the sky, although there are a few \FeII~knots visible in the central, blueshifted part of the flow (Figure~\ref{fig:feii_series_pyspeckit}).

The 1.26 \mum~ and 1.64 \mum~transitions can be used to measure extinction because they both originate from the same upper level ($^{4}D$) and have an intrinsic flux ratio F($\lambda$12567)/F($\lambda$16435) = 1.49 \citep{Smith2006}, however, the 1.64 \mum~line is coincident with an airglow line and provides uncertain fluxes. The 1.26 \mum~and 1.64 \mum~transitions are the highest S/N of the \FeII~lines, with all other detected lines approximately an order of magnitude fainter. Table~\ref{table:FeII_fluxes} lists the observed \FeII~fluxes (1.26 \mum, 1.29 \mum, 1.59 \mum, 1.64 \mum, and 1.66 \mum) and intensity-weighted radial velocities for various knots marked in Figure~\ref{fig:regions}. We subtracted the airglow contamination for the 1.64 \mum~fluxes in Table~\ref{table:FeII_fluxes} by estimating the airglow contamination using the same aperture on a nearby \FeII-free part of the sky and subtracting. We note that the radial velocity measurements for \FeII~have uncertainties of $\pm$~40 \kms.  Unlike for the \HH~radial velocities, we chose to report the radial velocity not as the median value among the observed transitions but as the value for the highest S/N transition (1.26 \mum); there was significant disagreement in the radial velocity measurements among the \FeII~transitions.

\subsection{Ionization Front and PDR} \label{sec:IonizationFront}

We detect emission from O, Fe$^{+}$, and P$^{+}$ (singly-ionized phosphorus) originating from the ionization front and PDR northeast of the Trapezium cluster, between the foreground M42 \HII~region and the background Orion A molecular cloud and OMC1 cloud core.   These atomic transitions are excited by far-UV radiation from the foreground Trapezium stars at the surface of the molecular cloud in which the BN/KL outflow is embedded.  Specifically, we detect in emission the 1.3164 \mum~\OI~line \citep{Marconi1998, Smith2006}, the \FeII~ lines at 1.256 \mum~ and 1.644 \mum~ and their related transitions, and 1.18877 \mum~\PII~\citep{Rudy1991,Rudy2001}. Except for \OI~and \FeII~at 1.256\mum, the lines are low S/N and integrated intensity maps could not be made.

With the PDR tracers, we detect an ionized feature oriented in the east-west direction at $\delta$ = -5:22:20 called the ``E-W Bright Bar" by \cite{Garcia-Diaz2007}.  At $\delta$ = -5:23:12 is the ``Trapezium Compact Bar", also in the east-west direction \citep{Garcia-Diaz2007}, and the Trapezium cluster resides in the 20\arcsec~missing box of data to the east at $\alpha$ = 5:35:16.  The atomic emission associated with these features is redshifted, $v_{\rm LSR}$ = +15--30 \kms.

These PDRs have been detected in many other tracers including recent velocity-resolved C$^{+}$ observations with $Herschel$ \citep{Goicoechea2015a} that detect the large ``East PDR" of which the E-W Bright Bar is only a small, western section.  \cite{Goicoechea2015a} detects C$^{+}$ at $v_{\rm LSR}$ = +9.5 \kms~with a typical line-width of 4-5 \kms, consistent within errors with this paper's radial velocity measurement ($v_{\rm LSR}$ = +15--30 \kms).  

\subsection{Source I and BN Reflection Nebulae} \label{sec:Continuum}

\begin{figure*}
     \begin{center}

        \subfigure{

            \includegraphics[width=\textwidth]{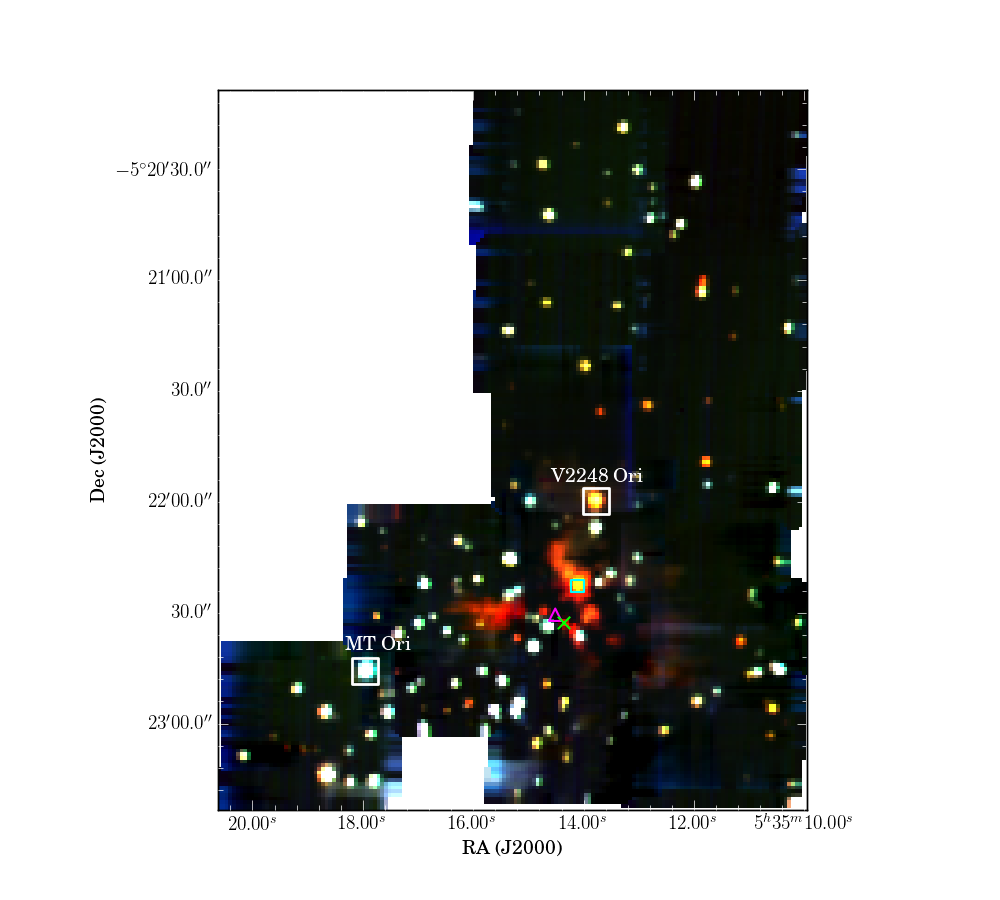}
     }

    \end{center}
    \caption{
        K (red), H (green), and J (blue) band continuum image that excludes all airglow lines and emission lines.  The cyan square, magenta triangle, and green X mark the locations of the BN object, source I, and source n, respectively. The two white squares show V2248 Ori and MT Ori. Proper motions show these three objects were within several hundred AU approximately 500 years ago.  
     }
   \label{fig:RGB_continuum}
\end{figure*}

Polarization studies by \cite{Hough1986} and \cite{Burton1991} showed that there is a diffuse \HH~reflection nebula surrounding the BN/KL outflow produced by the \HH~Peaks 1 and 2 \citep{Beckwith1978}, and continuum reflection nebulae caused mainly by source I.  The J, H, and K band continuum image (free of emission lines) in Figure~\ref{fig:RGB_continuum} shows the reflection nebulae surrounding source I and the BN object.  Polarization studies have confirmed the K band continuum is produced by scattered light from the young protostars source I and BN off aligned dust grains (e.g., \citealt{Hough1986}).  The J and H band continuum is dominated by scattered light from the bluer Trapezium stars, but we lack the sensitivity to detect this.  

\begin{figure*}
     \begin{center}

        \subfigure{

            \includegraphics[width=\textwidth]{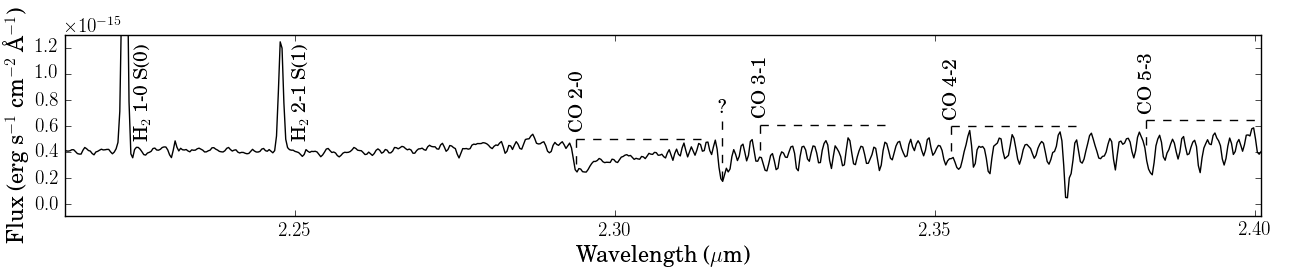}

       }

    \end{center}
    \caption{
        K-band spectrum integrated over the extent of source I's reflection nebula at (5:35:15.676, -05:22:29.60) with a 10.7\arcsec~$\times$~2.7\arcsec~aperture at PA=5\,$^{\circ}$.  The same aperture to the northeast at (5:35:17.246, -05:22:17.41) was used to subtract the nebula and airglow contamination.
     }
   \label{fig:sourceI_reflectionnebula_spectrum}
\end{figure*}

We extracted a spectrum of source I's reflection nebula to the east, using a larger aperture than \cite{Morino1998} and \cite{Testi2010}, and confirmed the presence of CO bandheads in absorption at $\lambda$~\textgreater~2.29 \mum~(Figure~\ref{fig:sourceI_reflectionnebula_spectrum}). There is an unidentified absorption line at 2.317 \mum~that could be an artifact, because it is coincident with an airglow line. CO absorption indicates a cool photosphere for source I (T \textless~5500 K), but could also be dominated by the cool ``photosphere" of a circumstellar disk \citep{Morino1998,Testi2010}.  Our spectra do not reach sufficient depth to detect any other photospheric lines. Because source I and the BN object are widely accepted as the sources of the BN/KL outflow, characterizing them is essential for distinguishing between the current models of the outflow's origin (e.g., \citealt{Tan2004}, \citealt{Bally2005}, and \citealt{Bally2015}).  

The spectrum of BN's reflection nebula exhibits \BrG~and \HeI~(2.05 \mum) absorption, confirming BN's consistency with a late-O/early-B type star \citep{Scoville1983,Hanson1996}, but is not of sufficient depth to detect other features.

\subsection{Jets and Disks} \label{sec:Jets}

Several Herbig Haro objects and other YSO jets are visible in \HH, \HI, \HeI, and \FeII~and are discussed further in this section.

\subsubsection{HH 201} \label{sec:HH201}

HH 201 (5:35:11.393, -5:21:53.97) is a well-studied Herbig Haro object of the same dynamical origin as other fingers in the BN/KL outflow \citep{Doi2004}.  It is the brightest of the knots in both the optical \citep{Munch1974} and near-IR \FeII~\citep{Bally2015}, but has only faint \HH~emission \citep{Graham2003} indicating that HH 201 lies in the foreground PDR.  

The peak \FeII~emission occurs at the fingertip (the northwest end) and is blueshifted ($v_{\rm LSR}$ = -80 \kms), indicating this knot is moving out of the OMC1 cloud core into the PDR. In the optical (H$\alpha$~and [S\,\textsc{ii}]), HH 201's radial velocity is about -270 \kms~\citep{Doi2004,Graham2003}, and this value agrees with the velocity \FeII~and Pa$\beta$~first appear in the PPV cube. It is possible that HH 201 is the brightest \FeII~source because iron's gas phase abundance is increased in the PDR (Section~\ref{sec:FeIIfingertips}).  

\begin{figure*}
     \begin{center}

        \subfigure{
            \includegraphics[width=\textwidth]{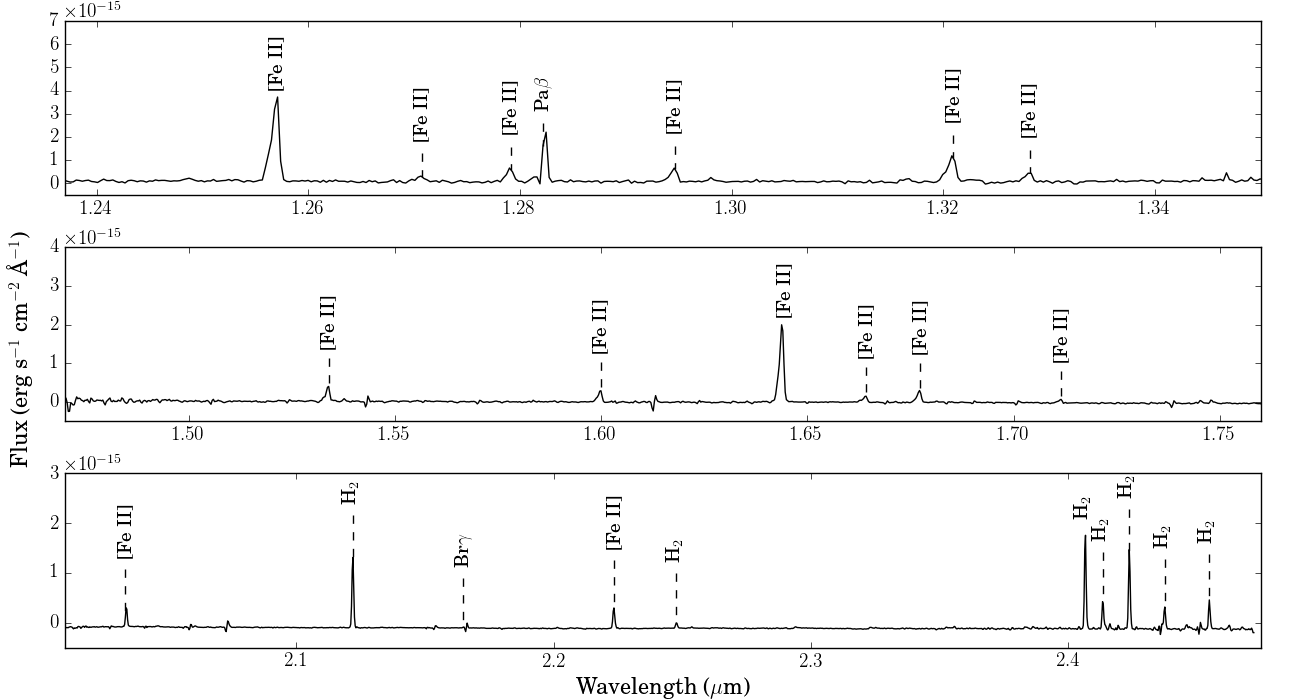}
        }

    \end{center}
    \caption{
        HH 201's spectrum integrated over a 6.75\arcsec~$\times$~3\arcsec~aperture with PA=330\,$^{\circ}$ at (5:35:11.5, -5:21:54.8).  The same aperture 8\arcsec~north was used to subtract the nebula and airglow emission.  The strong \HH~lines present here are probably confused from nearby regions, and the unmarked lines are residual airglow.
     }
   \label{fig:HH201_spectrum}
\end{figure*}

HH 201's \FeII~line profiles show an asymmetric line shape with a blue tail (Figure~\ref{fig:HH201_spectrum}).  \cite{Doi2004} showed HH 201 is a superposition of two bowshocks, explaining the asymmetry of the line profiles.  The ratio of HH 201's proper motion $\sim$~170 \kms~\citep{Hu1996} to its radial velocity suggests that HH 201 is inclined $\sim$ 60\,$^{\circ}$ to our line-of-sight, assuming the tilt of the knot away from the plane of the sky is given by $\theta_{jet}$ = tan$^{-1}$(V$_r$/V$_{PM}$), where V$_r$ is the average radial velocity and V$_{PM}$ is the proper motion.

In Figure~\ref{fig:regions}, HH 210 is marked as ``\FeII--10" in the left panel and ``10" in the right panel. It slightly overlaps with the \HH~feature \#25-W, which has the same position angle and is located to the southeast. \#25-W is redshifted ($v_{\rm LSR}$ = +22 \kms), in stark contrast to HH 210. \#25-W has a visual extinction 2-3 magnitudes greater than its surroundings, while HH 210's extinction does not stand out (Figure~\ref{fig:AV_polar}). Because HH 201's source is thought to be the same as the BN/KL outflow and the two are not bright in the same tracers, it is unlikely \#25-W and HH 201 are associated.

\subsubsection{HH 202} \label{sec:HH202}

\begin{figure*}
     \begin{center}

        \subfigure{
            \includegraphics[width=\textwidth]{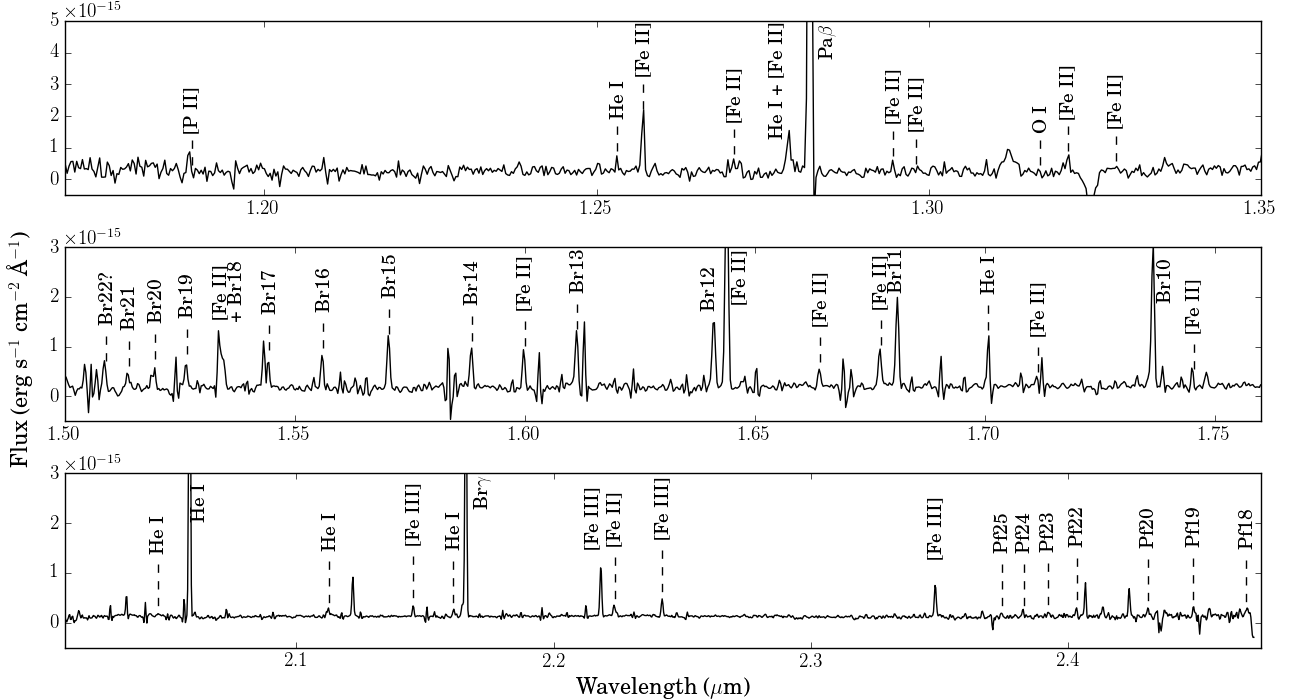}
        }

    \end{center}
    \caption{
        HH 202's spectrum integrated over two apertures: a 1.9\arcsec~$\times$~3.5\arcsec~ellipse with PA=40\,$^{\circ}$ at (05:35:11.7,-5:22:49.5) and a 2.3\arcsec~$\times$~3.2\arcsec ellipse with PA=330\,$^{\circ}$ at (05:35:11.7,-5:22:55.4).  The same pair of apertures offset 17\arcsec~to the south was used to subtract nebular and airglow contamination. Unmarked lines are residual airglow.
     }
   \label{fig:HH202}
\end{figure*}

HH 202 (5:35:11.640, -5:22:55.25) is a fully-ionized jet not associated dynamically with the BN/KL outflow.  Its source is unknown, although it may share a common origin with other nearby Herbig Haro objects \citep{ODell1997}. Photoionization by the Trapezium cluster is its primary excitation mechanism with negligible shock excitation \citep{ODell1997} and significant departures from Case B recombination \citep{Mesa-Delgado2009a}. HH 202 shows extended emission in many tracers including \HI~Pa$\beta$, the Brackett and Pfund series, \HeI, \FeII, \OI, and \FeIII~(Figure~\ref{fig:HH202}).

\subsubsection{HH 210} \label{sec:HH210}

\begin{figure*}
     \begin{center}

        \subfigure{
            \includegraphics[width=\textwidth]{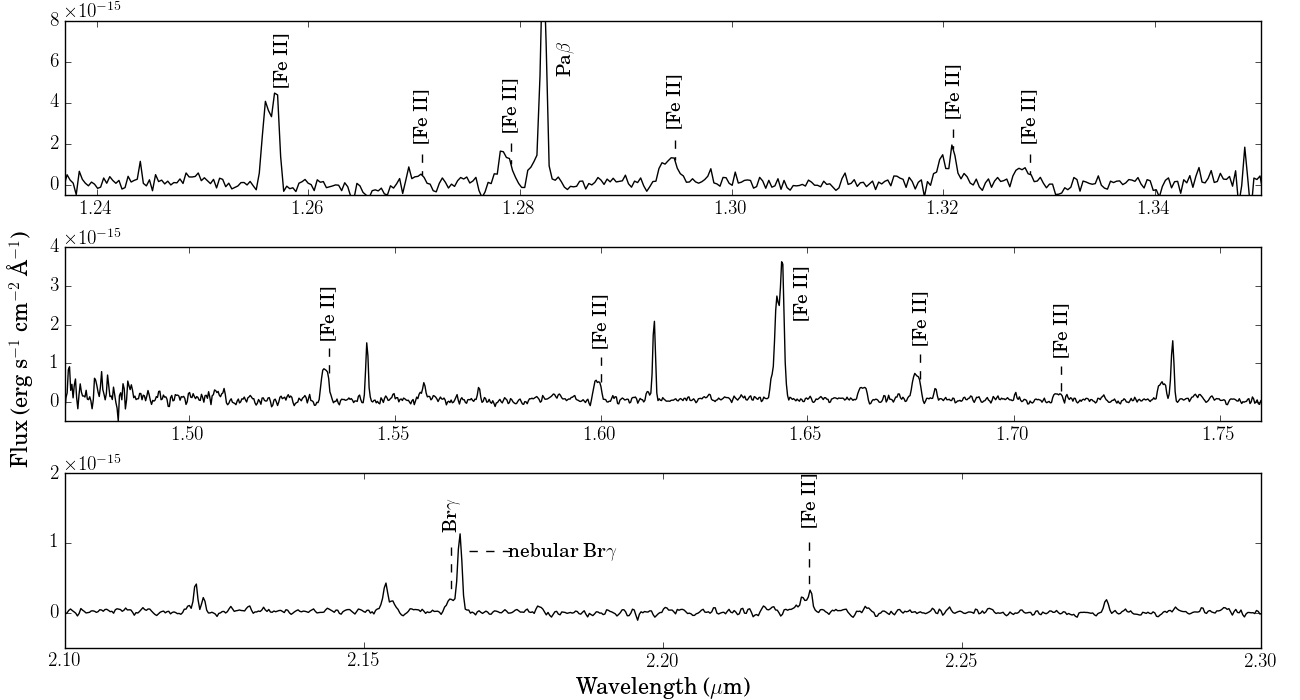}
        }

    \end{center}
    \caption{
        HH 210's spectrum integrated over a 0.84\arcsec~circular aperture at (5:35:15.482, -5:20:39.28).  The same aperture at (5:35:15.591, -05:20:13.22) was used to subtract nebular and airglow contamination. Unmarked lines are residual airglow.
     }
   \label{fig:HH210_spectrum}
\end{figure*}

\begin{figure*}
     \begin{center}

        \subfigure{
            \includegraphics[width=0.85\textwidth]{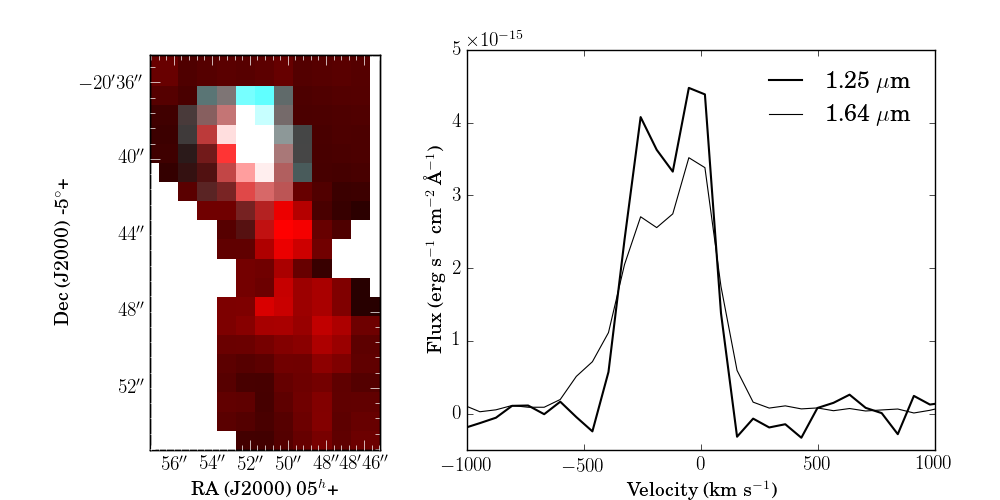}
        }

    \end{center}
    \caption{
        ($Left$) \FeII~1.26 \mum~integrated intensity map of HH 210.  Red shows emission with velocities between -130 and 150 \kms, and the blue and green layers are equally scaled and weighted to show emission with velocities between -470 and -130 \kms. ($Right$) HH 210's double-peaked \FeII~line profile. The thick line shows the 1.26 \mum~$^{6}D$ -- $^{4}D$ line and the thin line shows the 1.64 \mum~$^{6}D$ -- $^{4}$F line.  The peaks are at -260 \kms~and -30 \kms.  The 1.26 \mum~line is not contaminated by airglow, while the 1.64 \mum~line shows airglow coincident with redshifted peak, and the small bump blueward of the -260 \kms~peak is airglow.  
     }
   \label{fig:HH210_doublepeaked_FeII_profile}
\end{figure*}

HH 210 is a knot at (5:35:15.430, -5:20:39.95) of the BN/KL outflow and is only detected in \FeII~and \HI.  It is the only known Herbig Haro object associated with soft X-ray emission (COUP 703 and COUP 704; \citealt{Grosso2006}).  HH 210 has the highest proper motion of all the BN/KL fingers (309--425 \kms; \citealt{Doi2002}), and along with HH 201 (Section~\ref{sec:HH201}), is among the few visible at optical wavelengths as it lies in front of the dense gas in which the rest of the fingers are embedded. HH 210 is visible in almost all \FeII~lines in J and H bands, \HI~Br$\gamma$ and Pa$\beta$. We detect two kinematically distinct components in \FeII~emission that peak at $v_{\rm LSR}$ = -260 \kms~and -30 \kms~(Figure~\ref{fig:HH210_doublepeaked_FeII_profile}). The lack of \HH~emission also indicates that HH 210 lies in the PDR, agreeing with its blueshifted emission.

Like HH 201, the PDR's presence might explain HH 210's high \FeII~intensity, and its double peaked line profile and dual soft X-ray counterparts suggest that it is the superposition of two bowshocks.  Combined with \cite{Doi2002}'s proper motion measurement of 309--425 \kms, the radial velocities suggest the two bowshocks' full velocity vectors are separated by $\sim$20\,$^{\circ}$. The blueshifted component appears less spatially extended than the slower-moving redshifted component (Figure~\ref{fig:HH210_doublepeaked_FeII_profile}) possibly because Fe$^{\rm +}$ is ionized to Fe$^{\rm ++}$ so the \FeII~emission measure is smaller.

\subsubsection{OMC-1n} \label{sec:OMC1n}

\begin{figure*}
     \begin{center}

        \subfigure{

            \includegraphics[width=0.75\textwidth]{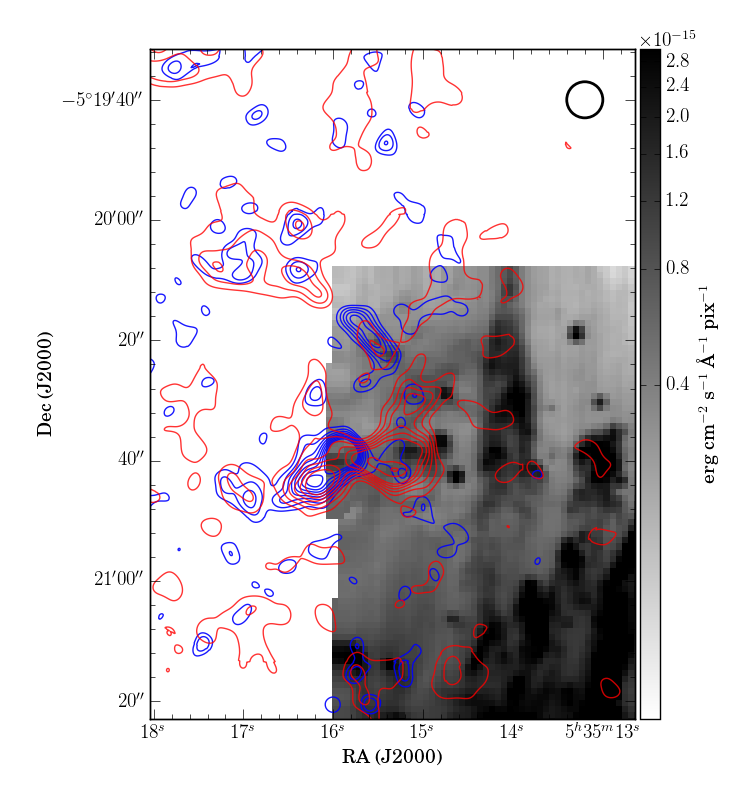}
     }

    \end{center}
    \caption{
        OMC-1n jets visible in the northeast corner of our PPV cube. Shown in grayscale is a single channel from the K-band PPV cube at $\lambda$=21219.7 \AA~($\Delta \lambda$=2.88 \AA). We do not show the \HH~1-0 S(1) integrated intensity map, because the S/N is low in the northeast corner of the map, so many of the Gaussian fits to the 1-0 S(1) line were not robust. The contours show CO (2-1) emission from \cite{Teixeira2016}. The red contours show emission with velocities from 0 to +40 \kms, with contour levels ranging from 14 to 140 Jy beam$^{-1}$ \kms~in steps of 14 Jy beam$^{-1}$ \kms. The blue contours show emission with velocities from -20 to 0 \kms, with contour levels ranging from 9 to 30 Jy beam$^{-1}$ \kms~in steps of 3 Jy beam$^{-1}$ \kms. The SMA's 3\arcsec~synthesized beam is shown in black in the upper right corner.
     }
   \label{fig:OMC1n}
\end{figure*}

A CO jet emitted from a protostar associated with the OMC-1n dense filaments may be visible in \HH~in the northeast corner of the PPV cube. Figure~\ref{fig:OMC1n} shows the overlap of CO (2-1) emission detected with the Submillimeter Array (SMA; \citealt{Teixeira2016}) and lobes of near-IR \HH~emission that do not appear kinematically associated with the BN/KL outflow. In this $\sim$40\arcsec~region around (5:35:15.5, -5:20:40), the CO and \HH~emission do not appear to coincide perfectly, but given \cite{Teixeira2016}'s 3\arcsec~pointing-accuracy estimate and our 1\arcsec-accurate WCS solution (which corresponds to 1 pixel accuracy given our 1\arcsec~pix$^{-1}$ pixel scale), overlap in the plane-of-the-sky is reasonable. They also coincide in velocity space. The \HH~double-lobe seen in Figure~\ref{fig:OMC1n} around (5:35:15, -5:20:40) appears over $v_{\rm LSR}$ = +22--34 \kms, and the redshifted CO emission appears over $v_{\rm LSR}$ = 0--40 \kms. South of the \HH~double-lobe is a faint circle of \HH~emission with radius 3.75\arcsec~(1500 AU at 414 pc), which may be the walls of a cavity evacuated by the jet.

\subsubsection{V2270 Ori} \label{sec:HH202}

\begin{figure}
     \begin{center}

        \subfigure{
            \includegraphics[width=0.5\textwidth]{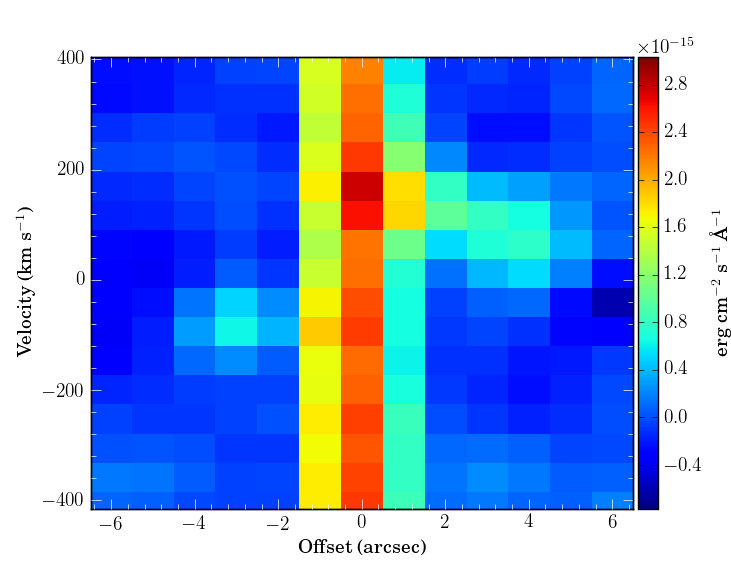}
        }

    \end{center}
    \caption{
        PV diagram of the V2270 Ori \FeII~(1.64 \mum) bipolar jet. The position axis (labeled ``Offset") is centered on V2270 Ori, and the velocity axis is centered on the \FeII~line's vacuum rest wavelength (1.6439 \mum). The blueshifted jet is centered around -100 \kms, and the redshifted jet appears to have a velocity gradient extending from approximately +125 \kms~near the star, to +65 \kms~away from the star. The uncertainties in these velocities are large.
     }
   \label{fig:pv_V2270Ori}
\end{figure}

The V2270 Ori (05:35:15.394, -05:21:14.11) bipolar outflow detected in the adaptive optics \FeII~image of \cite{Bally2015} is also detected in this dataset (Figure~\ref{fig:pv_V2270Ori}). The southwestern part of the jet is redshifted, and the northeastern jet is blueshifted. When applying our Gaussian fitting routine to this region, we find for the redshifted jet intensity-weighted $v_{\rm LSR}$ $\approx$~+100 \kms~with respect to the rest frame of the Orion Nebula \FeII~emission, and $v_{\rm LSR}$ $\approx$~-25 \kms~for the blueshifted jet. Using a position-velocity (PV) diagram, we find that the redshifted jet shows a negative velocity gradient extending from approximately +125 \kms~near the star, to +65 \kms~away from the star. The blueshifted jet is centered around -100 \kms. Because V2270 Ori's jet has low S/N in our observations, the velocity centroid measurements are large.

\begin{deluxetable}{lllr}
\tablecolumns{4}
\tablewidth{0pt}
\tabletypesize{\scriptsize}
\tablecaption{Coordinates for labeled \HH~features in Figure~\ref{fig:regions} \label{table:H2_coords}}
\tablehead{\colhead{Number} & 
                   \colhead{RA} & 
                   \colhead{Dec} &
                   \colhead{$v_{\rm LSR}^*$} \\
                   \colhead{} & 
                   \colhead{(J2000)} & 
                   \colhead{(J2000)} &
                   \colhead{(\kms)} 
                   }
\startdata
1 & 5:35:14.816 & -5:20:36.99 & 20 $\pm$~5 \\
2 & 5:35:14.004 & -5:20:30.57 & 9 $\pm$~12 \\
3 & 5:35:13.051 & -5:20:39.44 & 10 $\pm$~4 \\
4 & 5:35:12.209 & -5:20:39.76 & -2 $\pm$~28 \\
5 & 5:35:15.721 & -5:21:13.3 & 10 $\pm$~5 \\
6 & 5:35:13.369 & -5:20:55.58 & 7 $\pm$~5 \\
7 & 5:35:11.597 & -5:20:54.32 & -19 $\pm$~27 \\
8 & 5:35:11.864 & -5:21:17.11 & 4 $\pm$~4 \\
9 & 5:35:10.548 & -5:21:28.19 & 11 $\pm$~6 \\
10 & 5:35:12.002 & -5:21:42.85 & 2 $\pm$~7 \\
11 & 5:35:13.72 & -5:21:52.94 & 1 $\pm$~8 \\
12 & 5:35:14.739 & -5:21:56.83 & 4 $\pm$~5 \\
13 & 5:35:16.845 & -5:22:8.69 & 4 $\pm$~4 \\
14 & 5:35:14.963 & -5:22:21.62 & 1 $\pm$~5 \\
15 & 5:35:11.313 & -5:22:19.14 & 11 $\pm$~4 \\
16 & 5:35:14.645 & -5:22:32.61 & -15 $\pm$~8 \\
17 & 5:35:14.068 & -5:22:37.82 & -8 $\pm$~8 \\
18 & 5:35:13.703 & -5:22:39.2 & -11 $\pm$~6 \\
19 & 5:35:12.818 & -5:22:47.94 & 7 $\pm$~1 \\
20 & 5:35:11.849 & -5:22:38.98 & 12 $\pm$~3 \\
21 & 5:35:15.345 & -5:22:37.18 & 19 $\pm$~4 \\
22 & 5:35:17.947 & -5:23:6.31 & 2 $\pm$~2 \\
23 & 5:35:18.795 & -5:23:9.79 & 8 $\pm$~2 \\
24 & 5:35:19.367 & -5:23:12.02 & 7 $\pm$~4 \\
25 & 5:35:13.109 & -5:22:3.84 & 8 $\pm$~8 \\
26 & 5:35:13.576 & -5:22:3.57 & 1 $\pm$~8 \\
27 & 5:35:13.205 & -5:22:21.79 & -4 $\pm$~10 \\
\enddata
\tablenotetext{*}{ \HH~intensity-weighted radial velocities measured as the median value of the measurements for the transitions reported in Table~\ref{table:H2_fluxes}.  The uncertainty is reported in this table is the standard deviation of the measurements.}
\end{deluxetable}

\section{Summary} \label{sec:Summary}

We have presented near-IR (1.1 -- 2.4 \mum) PPV cubes of the Orion BN/KL outflow and integrated intensity and kinematic maps in 13 \HH~and 4 \FeII~lines.  The kinematic maps have clarified the velocity structure of the 500-year-old, wide-angle outflow. In the central arcminute of the flow, the \HH~emission is dominated by blueshifted fingers and knots on the near-side of the flow.  When combined with proper motion measurements from \cite{Bally2015}, we see that most of the outer fingers' motion lies predominantly in the plane of the sky, save for a few knots and fingers that exhibit significant redshift.  The redshifted features correspond to regions of low extinction, indicating that they are on the far side of the flow and are seen through a cavity in the dense gas.  

We also present tables of \HH~and \FeII~flux ratios for a sample of features and a pixel-by-pixel flux ratio map for the \HH~1-0 S(1) and 2-1 S(1) transitions.  The 1-0 S(1)/2-1 S(1) line ratios are consistent with shock excitation, although in the outer regions of the flow, the ratios indicate that UV excitation is present but not dominant.  These line ratios (presented as observed and dereddened) can be used for modeling with shock or PDR codes to derive the true excitation conditions of the BN/KL outflow.

Present in the PPV cubes are several bright YSO jets that we highlight: HH 201, HH 202, HH 210, V2270 Ori, and two protostellar jets north of HH 210.  We also detect two ionization fronts and PDRs and the reflection nebulae surrounding source I and the BN object.

\renewcommand{\labelitemi}{--}

\acknowledgments
Observations were obtained with the Apache Point Observatory 3.5-meter telescope, which is owned and operated by the Astrophysical Research Consortium. We thank Paula Teixeira for providing the CO data cubes from the SMA of OMC1-n. AY acknowledges support from HST grant AR-13267-002 to the University of Colorado at Boulder during a portion of this work, and thanks Kevin France for helpful discussions. This research made use of the following Python packages:
\begin{itemize}
\item \texttt{aplpy} (http://aplpy.github.io)
\item \texttt{astropy} \citep{Robitaille2013}
\item \texttt{ipython} (http://ipython.org/ )
\item \texttt{molecular-hydrogen} (https://github.com/keflavich/molecular\_hydrogen)
\item \texttt{pvextractor} (http://pvextractor.readthedocs.org/)
\item \texttt{pyregion} (http://pyregion.readthedocs.org/) 
\item \texttt{pyspeckit} \citep{Ginsburg2011} 
\item \texttt{scicatalog} (https://github.com/parkus/scicatalog)
\item \texttt{sdpy} (https://github.com/keflavich/sdpy)
\item \texttt{spectral-cube} (http://spectral-cube.readthedocs.org/)
\item \texttt{wcsaxes} (wcsaxes.rtfd.org)
\end{itemize}

\bibliography{Orion_Tspec_1_arxiv.bbl}{}

\begin{thebibliography}{}
\expandafter\ifx\csname natexlab\endcsname\relax\def\natexlab#1{#1}\fi

\bibitem[{Bally {et~al.}(2011)Bally, Cunningham, Moeckel, Burton, Smith, Frank,
  \& Nordlund}]{Bally2011}
Bally, J., Cunningham, N.~J., Moeckel, N., {et~al.} 2011, The Astrophysical
  Journal, 727, 113

\bibitem[{Bally {et~al.}(2015)Bally, Ginsburg, Silvia, \&
  Youngblood}]{Bally2015}
Bally, J., Ginsburg, A., Silvia, D., \& Youngblood, A. 2015, Astronomy {\&}
  Astrophysics, 579, A130

\bibitem[{Bally \& Zinnecker(2005)}]{Bally2005}
Bally, J., \& Zinnecker, H. 2005, The Astronomical Journal, 129, 2281

\bibitem[{Beckwith {et~al.}(1978)Beckwith, Persson, Neugebauer, \&
  Becklin}]{Beckwith1978}
Beckwith, S., Persson, S.~E., Neugebauer, G., \& Becklin, E.~E. 1978, The
  Astrophysical Journal, 223, 464

\bibitem[{Black \& Dalgarno(1976)}]{Black1976}
Black, J.~H., \& Dalgarno, A. 1976, The Astrophysical Journal, 203, 132

\bibitem[{Black \& van Dishoeck(1987)}]{Black1987}
Black, J.~H., \& van Dishoeck, E.~F. 1987, The Astrophysical Journal, 322, 412

\bibitem[{Burton {et~al.}(1991)Burton, Minchin, Hough, Aspin, Axon, \&
  Bailey}]{Burton1991}
Burton, M.~G., Minchin, N.~R., Hough, J.~H., {et~al.} 1991, The Astrophysical
  Journal, 375, 611

\bibitem[{Cardelli {et~al.}(1989)Cardelli, Clayton, \& Mathis}]{Cardelli1989}
Cardelli, J.~A., Clayton, G.~C., \& Mathis, J.~S. 1989, The Astrophysical
  Journal, 345, 245

\bibitem[{Chatterjee \& Tan(2012)}]{Chatterjee2012}
Chatterjee, S., \& Tan, J.~C. 2012, The Astrophysical Journal, 754, 152

\bibitem[{Chen {et~al.}(2014)Chen, Goldsmith, Viti, Snell, Lis, Benz, Bergin,
  Black, Caselli, Encrenaz, Falgarone, Goicoechea, Hjalmarson, Hollenbach,
  Kaufman, Melnick, Neufeld, Pagani, van~der Tak, van Dishoeck, \&
  Yıldız}]{Chen2014}
Chen, J.-H., Goldsmith, P.~F., Viti, S., {et~al.} 2014, The Astrophysical
  Journal, 793, 111

\bibitem[{Colgan {et~al.}(2007)Colgan, Schultz, Kaufman, Erickson, \&
  Hollenbach}]{Colgan2007}
Colgan, S. W.~J., Schultz, A. S.~B., Kaufman, M.~J., Erickson, E.~F., \&
  Hollenbach, D.~J. 2007, The Astrophysical Journal, 671, 536

\bibitem[{Cushing {et~al.}(2004)Cushing, Vacca, \& Rayner}]{Cushing2004}
Cushing, M., Vacca, W., \& Rayner, J. 2004, Publications of the Astronomical
  Society of the Pacific, 116, 362

\bibitem[{Cyganowski {et~al.}(2008)Cyganowski, Whitney, Holden, Braden, Brogan,
  Churchwell, Indebetouw, Watson, Babler, Benjamin, Gomez, Meade, Povich,
  Robitaille, \& Watson}]{Cyganowski2008}
Cyganowski, C.~J., Whitney, B.~A., Holden, E., {et~al.} 2008, The Astronomical
  Journal, 136, 2391

\bibitem[{Doi {et~al.}(2004)Doi, O'Dell, \& Hartigan}]{Doi2004}
Doi, T., O'Dell, C.~R., \& Hartigan, P. 2004, The Astronomical Journal, 127,
  3456

\bibitem[{Doi {et~al.}(2002)Doi, O'Dell, \& Hartigan}]{Doi2002}
Doi, T., O'Dell, C.~R., \& Hartigan, P. 2002, The Astronomical Journal, 124,
  445

\bibitem[{Eisl{\"{o}}ffel(2000)}]{Eisloffel2000}
Eisl{\"{o}}ffel, Â. 2000, Astronomy and Astrophysics

\bibitem[{Garc{\'{\i}}a-D{\'{\i}}az \& Henney(2007)}]{Garcia-Diaz2007}
Garc{\'{\i}}a-D{\'{\i}}az, M.~T., \& Henney, W.~J. 2007, The Astronomical
  Journal, 133, 952

\bibitem[{Garc{\'{\i}}a-D{\'{\i}}az {et~al.}(2008)Garc{\'{\i}}a-D{\'{\i}}az,
  Henney, L{\'{o}}pez, \& Doi}]{Garcia-Diaz2008}
Garc{\'{\i}}a-D{\'{\i}}az, Â., Henney, Â., L{\'{o}}pez, Â., \& Doi, Â. 2008,
  Revista Mexicana de Astronom{\'{\i}}a y Astrof{\'{\i}}sica Vol. 44, 44, 181

\bibitem[{Geballe {et~al.}(1982)Geballe, Russell, \& Nadeau}]{Geballe1982}
Geballe, T.~R., Russell, R.~W., \& Nadeau, D. 1982, The Astrophysical Journal,
  259, L47

\bibitem[{Ginsburg \& Mirocha(2011)}]{Ginsburg2011}
Ginsburg, A., \& Mirocha, J. 2011, Astrophysics Source Code Library

\bibitem[{Goddi {et~al.}(2011)Goddi, Humphreys, Greenhill, Chandler, \&
  Matthews}]{Goddi2011}
Goddi, C., Humphreys, E. M.~L., Greenhill, L.~J., Chandler, C.~J., \& Matthews,
  L.~D. 2011, The Astrophysical Journal, 728, 15

\bibitem[{Goicoechea {et~al.}(2015{\natexlab{a}})Goicoechea, Chavarr{\'{\i}}a,
  Cernicharo, Neufeld, Vavrek, Bergin, Cuadrado, Encrenaz, Etxaluze, Melnick,
  \& Polehampton}]{Goicoechea2015}
Goicoechea, J.~R., Chavarr{\'{\i}}a, L., Cernicharo, J., {et~al.}
  2015{\natexlab{a}}, The Astrophysical Journal, 799, 102

\bibitem[{Goicoechea {et~al.}(2015{\natexlab{b}})Goicoechea, Teyssier,
  Etxaluze, Goldsmith, Ossenkopf, Gerin, Bergin, Black, Cernicharo, Cuadrado,
  Encrenaz, Falgarone, Fuente, Hacar, Lis, Marcelino, Melnick, Muller, Persson,
  Pety, Rollig, Schilke, Simon, Snell, \& Stutzki}]{Goicoechea2015a}
Goicoechea, Â., Teyssier, Â., Etxaluze, Â., {et~al.} 2015{\natexlab{b}}, eprint
  arXiv:1508.03801

\bibitem[{G{\'{o}}mez {et~al.}(2008)G{\'{o}}mez, Rodr{\'{\i}}guez, Loinard,
  Lizano, Allen, Poveda, \& Menten}]{Gomez2008}
G{\'{o}}mez, L., Rodr{\'{\i}}guez, L.~F., Loinard, L., {et~al.} 2008, The
  Astrophysical Journal, 685, 333

\bibitem[{Gomez {et~al.}(2005)Gomez, RodrIguez, Loinard, Lizano, Poveda, \&
  Allen}]{Gomez2005}
Gomez, L., RodrIguez, L.~F., Loinard, L., {et~al.} 2005, The Astrophysical
  Journal, 635, 1166

\bibitem[{Graham {et~al.}(2003)Graham, Meaburn, \& Redman}]{Graham2003}
Graham, M.~F., Meaburn, J., \& Redman, M.~P. 2003, Monthly Notices of the Royal
  Astronomical Society, 343, 419

\bibitem[{Greenhouse {et~al.}(1991)Greenhouse, Woodward, Thronson, Rudy,
  Rossano, Erwin, \& Puetter}]{Greenhouse1991}
Greenhouse, M.~A., Woodward, C.~E., Thronson, Harley~A., J., {et~al.} 1991, The
  Astrophysical Journal, 383, 164

\bibitem[{Grosso {et~al.}(2006)Grosso, Feigelson, Getman, Kastner, Bally, \&
  McCaughrean}]{Grosso2006}
Grosso, N., Feigelson, E.~D., Getman, K.~V., {et~al.} 2006, Astronomy and
  Astrophysics, 448, L29

\bibitem[{Gutermuth {et~al.}(2004)Gutermuth, Megeath, Muzerolle, Allen, Pipher,
  Myers, \& Fazio}]{Gutermuth2004}
Gutermuth, R.~A., Megeath, S.~T., Muzerolle, J., {et~al.} 2004, The
  Astrophysical Journal Supplement Series, 154, 374

\bibitem[{Hanson {et~al.}(1996)Hanson, Conti, \& Rieke}]{Hanson1996}
Hanson, M.~M., Conti, P.~S., \& Rieke, M.~J. 1996, The Astrophysical Journal
  Supplement Series, 107, 281

\bibitem[{Hough {et~al.}(1986)Hough, Axon, Burton, Gatley, Sato, Bailey,
  McCaughrean, McLean, Nagata, Allen, Garden, Hasegawa, Hayashi, Kaifu,
  Morimoto, \& Walther}]{Hough1986}
Hough, Â., Axon, Â., Burton, Â., {et~al.} 1986, Monthly Notices of the Royal
  Astronomical Society (ISSN 0035-8711), 222, 629

\bibitem[{Hu(1996)}]{Hu1996}
Hu, X. 1996, The Astronomical Journal, 112, 2712

\bibitem[{Lord(1992)}]{Lord1992}
Lord, Â. 1992, NASA Technical Memorandum 103957

\bibitem[{Marconi {et~al.}(1998)Marconi, Testi, Natta, \&
  Walmsley}]{Marconi1998}
Marconi, Â., Testi, Â., Natta, Â., \& Walmsley, Â. 1998, Astronomy and
  Astrophysics

\bibitem[{Markwardt(2009)}]{Markwardt2009}
Markwardt, Â. 2009, Astronomical Data Analysis Software and Systems XVIII ASP
  Conference Series, 411

\bibitem[{Menten {et~al.}(2007)Menten, Reid, Forbrich, \&
  Brunthaler}]{Menten2007}
Menten, K.~M., Reid, M.~J., Forbrich, J., \& Brunthaler, A. 2007, Astronomy and
  Astrophysics, 474, 515

\bibitem[{Mesa-Delgado {et~al.}(2009)Mesa-Delgado, Esteban,
  Garc{\'{\i}}a-Rojas, Luridiana, Bautista, Rodr{\'{\i}}guez,
  L{\'{o}}pez-Mart{\'{\i}}n, \& Peimbert}]{Mesa-Delgado2009a}
Mesa-Delgado, A., Esteban, C., Garc{\'{\i}}a-Rojas, J., {et~al.} 2009, Monthly
  Notices of the Royal Astronomical Society, 395, 855

\bibitem[{Morino {et~al.}(1998)Morino, Yamashita, Hasegawa, \&
  Nakano}]{Morino1998}
Morino, J.~I., Yamashita, T., Hasegawa, T., \& Nakano, T. 1998, Nature, 393,
  340

\bibitem[{Mouri \& Taniguchi(2000)}]{Mouri2000}
Mouri, H., \& Taniguchi, Y. 2000, The Astrophysical Journal, 534, L63

\bibitem[{M{\"{u}}nch \& Taylor(1974)}]{Munch1974}
M{\"{u}}nch, G., \& Taylor, K. 1974, The Astrophysical Journal, 192, L93

\bibitem[{O'Dell {et~al.}(1997)O'Dell, Hartigan, Lane, Wong, Burton, Raymond,
  \& Axon}]{ODell1997}
O'Dell, C.~R., Hartigan, P., Lane, W.~M., {et~al.} 1997, The Astronomical
  Journal, 114, 730

\bibitem[{Robitaille {et~al.}(2013)Robitaille, Tollerud, Greenfield,
  Droettboom, Bray, Aldcroft, Davis, Ginsburg, Price-Whelan, Kerzendorf,
  Conley, Crighton, Barbary, Muna, Ferguson, Grollier, Parikh, Nair,
  G{\"{u}}nther, Deil, Woillez, Conseil, Kramer, Turner, Singer, Fox, Weaver,
  Zabalza, Edwards, {Azalee Bostroem}, Burke, Casey, Crawford, Dencheva, Ely,
  Jenness, Labrie, Lim, Pierfederici, Pontzen, Ptak, Refsdal, Servillat, \&
  Streicher}]{Robitaille2013}
Robitaille, T.~P., Tollerud, E.~J., Greenfield, P., {et~al.} 2013, Astronomy
  {\&} Astrophysics, 558, A33

\bibitem[{Rodr{\'{\i}}guez {et~al.}(2005)Rodr{\'{\i}}guez, Poveda, Lizano, \&
  Allen}]{Rodriguez2005}
Rodr{\'{\i}}guez, L.~F., Poveda, A., Lizano, S., \& Allen, C. 2005, The
  Astrophysical Journal, 627, L65

\bibitem[{Rousselot {et~al.}(2000)Rousselot, Lidman, Cuby, Moreels, \&
  Monnet}]{Rousselot2000}
Rousselot, Â., Lidman, Â., Cuby, Â.-G., Moreels, Â., \& Monnet, Â. 2000,
  Astronomy and Astrophysics

\bibitem[{Rudy {et~al.}(2001)Rudy, Lynch, Mazuk, Puetter, \&
  Dearborn}]{Rudy2001}
Rudy, R.~J., Lynch, D.~K., Mazuk, S., Puetter, R.~C., \& Dearborn, D. S.~P.
  2001, The Astronomical Journal, 121, 362

\bibitem[{Rudy {et~al.}(1991)Rudy, Rossano, Erwin, \& Puetter}]{Rudy1991}
Rudy, R.~J., Rossano, G.~S., Erwin, P., \& Puetter, R.~C. 1991, The
  Astrophysical Journal, 368, 468

\bibitem[{Sahai {et~al.}(2008)Sahai, Claussen, {S{\'{a}}nchez Contreras},
  Morris, \& Sarkar}]{Sahai2008}
Sahai, R., Claussen, M., {S{\'{a}}nchez Contreras}, C., Morris, M., \& Sarkar,
  G. 2008, The Astrophysical Journal, 680, 483

\bibitem[{Salji {et~al.}(2015)Salji, Richer, Buckle, Hatchell, Kirk, Beaulieu,
  Berry, Broekhoven-Fiene, Currie, Fich, Jenness, Johnstone, Mottram, Nutter,
  Pattle, Pineda, Quinn, Tisi, Walker-Smith, Francesco, Hogerheijde,
  Ward-Thompson, Bastien, Butner, Chen, Chrysostomou, Coude, Davis,
  Drabek-Maunder, Duarte-Cabral, Fiege, Friberg, Friesen, Fuller, Graves,
  Greaves, Gregson, Holland, Joncas, Kirk, Knee, Mairs, Marsh, Matthews,
  Moriarty-Schieven, Rawlings, Robertson, Rosolowsky, Rumble, Sadavoy, Thomas,
  Tothill, Viti, White, Wilson, Wouterloot, Yates, \& Zhu}]{Salji2015}
Salji, C.~J., Richer, J.~S., Buckle, J.~V., {et~al.} 2015, Monthly Notices of
  the Royal Astronomical Society, 449, 1769

\bibitem[{Scoville {et~al.}(1983)Scoville, Kleinmann, Hall, \&
  Ridgway}]{Scoville1983}
Scoville, N., Kleinmann, S.~G., Hall, D. N.~B., \& Ridgway, S.~T. 1983, The
  Astrophysical Journal, 275, 201

\bibitem[{Shull(1978)}]{Shull1978}
Shull, J.~M. 1978, The Astrophysical Journal, 219, 877

\bibitem[{Smith \& Hartigan(2006)}]{Smith2006}
Smith, N., \& Hartigan, P. 2006, The Astrophysical Journal, 638, 1045

\bibitem[{Smith {et~al.}(2009)Smith, Whitney, Conti, {De Pree}, \&
  Jackson}]{Smith2009}
Smith, N., Whitney, B.~A., Conti, P.~S., {De Pree}, C.~G., \& Jackson, J.~M.
  2009, Monthly Notices of the Royal Astronomical Society, 399, 952

\bibitem[{Snell {et~al.}(1984)Snell, Scoville, Sanders, \&
  Erickson}]{Snell1984}
Snell, R.~L., Scoville, N.~Z., Sanders, D.~B., \& Erickson, N.~R. 1984, The
  Astrophysical Journal, 284, 176

\bibitem[{Tan(2004)}]{Tan2004}
Tan, J.~C. 2004, The Astrophysical Journal, 607, L47

\bibitem[{Teixeira {et~al.}(2016)Teixeira, Takahashi, Zapata, \&
  Ho}]{Teixeira2016}
Teixeira, P.~S., Takahashi, S., Zapata, L.~A., \& Ho, P. T.~P. 2016, Astronomy
  {\&} Astrophysics, 587, A47

\bibitem[{Testi {et~al.}(2010)Testi, Tan, \& Palla}]{Testi2010}
Testi, L., Tan, J.~C., \& Palla, F. 2010, Astronomy {\&} Astrophysics, 522, A44

\bibitem[{Vacca {et~al.}(2003)Vacca, Cushing, \& Rayner}]{Vacca2003}
Vacca, W., Cushing, M., \& Rayner, J. 2003, Publications of the Astronomical
  Society of the Pacific, 115, 389

\bibitem[{Wolfire \& Konigl(1991)}]{Wolfire1991}
Wolfire, M.~G., \& Konigl, A. 1991, The Astrophysical Journal, 383, 205

\bibitem[{Zapata {et~al.}(2009)Zapata, Schmid-Burgk, Ho, Rodr{\'{\i}}guez, \&
  Menten}]{Zapata2009}
Zapata, L.~A., Schmid-Burgk, J., Ho, P. T.~P., Rodr{\'{\i}}guez, L.~F., \&
  Menten, K.~M. 2009, The Astrophysical Journal, 704, L45

\bibitem[{Zapata {et~al.}(2013)Zapata, Schmid-Burgk, P{\'{e}}rez-Goytia, Ho,
  Rodr{\'{\i}}guez, Loinard, \& Cruz-Gonz{\'{a}}lez}]{Zapata2013}
Zapata, L.~A., Schmid-Burgk, J., P{\'{e}}rez-Goytia, N., {et~al.} 2013, The
  Astrophysical Journal, 765, L29

\end{thebibliography}
\bibliographystyle{apj}


\clearpage

\begin{turnpage}
\begin{deluxetable}{lccccccccccccr}
\tablecolumns{14}
\tablewidth{0pt}
\tabletypesize{\scriptsize}
\tablecaption{Observed \HH~fluxes for features labeled in Figure~\ref{fig:regions} \label{table:H2_fluxes}}
\tablehead{\colhead{No.} & 
                  \colhead{1-0 S(1)\tablenotemark{a,b}} & 
                  \colhead{1-0 S(0)\tablenotemark{c}} & 
                  \colhead{1-0 S(7)} & 
                  \colhead{1-0 S(8)} & 
                  \colhead{1-0 S(9)} & 
                  \colhead{2-1 S(0)} & 
                  \colhead{2-1 S(1)} & 
                  \colhead{3-2 S(3)} & 
                  \colhead{3-2 S(5)} & 
                  \colhead{1-0 Q(1)} & 
                  \colhead{1-0 Q(2)} & 
                  \colhead{1-0 Q(3)} & 
                  \colhead{1-0 Q(4)} 
                  }
\startdata
1 & 6.20$\times$10$^{-15}$ & 0.27 & 0.18 & -- & -- & -- & 0.09 & -- & -- & 1.48 & -- & 0.99 & -- \\
2 & 5.31$\times$10$^{-14}$ & -- & 0.32 & -- & 0.14 & -- & -- & -- & -- & -- & -- & 1.26 & -- \\
3 & 7.71$\times$10$^{-14}$ & -- & -- & -- & -- & -- & -- & -- & -- & 1.53 & -- & 1.19 & -- \\
4 & 5.71$\times$10$^{-14}$ & 0.27 & 0.29 & -- & -- & -- & 0.15 & -- & -- & 1.55 & -- & 1.26 & 0.43 \\
5 & 3.33$\times$10$^{-14}$ & -- & 0.15 & -- & -- & -- & -- & -- & -- & 2.05 & -- & 1.56 & -- \\
6 & 2.05$\times$10$^{-14}$ & 0.26 & 0.25 & -- & -- & -- & 0.12 & -- & -- & 1.38 & -- & 1.18 & -- \\
7 & 2.77$\times$10$^{-14}$ & 0.26 & -- & -- & -- & -- & -- & -- & -- & 1.56 & -- & 1.24 & 0.36 \\
8 & 1.02$\times$10$^{-13}$ & -- & 0.25 & -- & -- & -- & 0.12 & -- & -- & 1.29 & -- & 1.0 & -- \\
9 & 5.28$\times$10$^{-14}$ & -- & -- & -- & -- & -- & -- & -- & -- & -- & -- & 1.26 & -- \\
10 & 9.16$\times$10$^{-14}$ & 0.26 & 0.17 & -- & 0.04 & -- & 0.09 & -- & -- & 1.26 & 0.36 & 1.03 & 0.22 \\
11 & 4.59$\times$10$^{-13}$ & 0.27 & 0.14 & 0.02 & 0.03 & 0.03 & 0.11 & 0.02 & 0.01 & 1.25 & 0.37 & 1.11 & 0.31 \\
12 & 1.22$\times$10$^{-13}$ & 0.29 & 0.1 & -- & -- & 0.03 & 0.11 & 0.02 & -- & 1.52 & 0.41 & 1.28 & 0.33 \\
13 & 1.93$\times$10$^{-14}$ & 0.29 & -- & -- & -- & -- & 0.13 & -- & -- & 2.04 & -- & 1.41 & -- \\
14 & 7.27$\times$10$^{-14}$ & 0.28 & 0.12 & -- & -- & -- & 0.1 & -- & -- & 1.5 & -- & 1.18 & 0.2 \\
15 & 1.39$\times$10$^{-13}$ & 0.27 & 0.17 & -- & -- & -- & -- & -- & -- & -- & -- & -- & -- \\
16 & 1.81$\times$10$^{-13}$ & 0.28 & 0.13 & -- & 0.02 & 0.03 & 0.09 & 0.02 & -- & 1.29 & 0.36 & 1.14 & 0.32 \\
17 & 2.56$\times$10$^{-13}$ & 0.28 & 0.13 & -- & 0.02 & -- & 0.08 & 0.02 & -- & 1.35 & 0.36 & 1.14 & 0.3 \\
18 & 9.33$\times$10$^{-14}$ & 0.28 & 0.11 & -- & 0.02 & 0.03 & 0.09 & 0.01 & -- & 1.34 & 0.39 & 1.25 & 0.33 \\
19 & 2.43$\times$10$^{-13}$ & 0.3 & 0.09 & -- & -- & -- & 0.11 & -- & -- & 1.72 & -- & 1.47 & 0.36 \\
20 & 1.07$\times$10$^{-13}$ & 0.27 & 0.14 & -- & -- & -- & 0.11 & -- & -- & 1.07 & -- & 0.96 & 0.27 \\
21 & 2.26$\times$10$^{-13}$ & 0.28 & 0.1 & -- & -- & -- & 0.09 & -- & -- & 1.2 & 0.43 & 0.95 & 0.3 \\
22 & 5.11$\times$10$^{-14}$ & 0.29 & 0.21 & -- & -- & -- & 0.12 & -- & -- & -- & -- & -- & -- \\
23 & 1.51$\times$10$^{-14}$ & 0.3 & 0.17 & -- & -- & -- & 0.14 & -- & -- & -- & -- & 0.8 & -- \\
24 & 3.85$\times$10$^{-14}$ & 0.28 & 0.2 & -- & -- & -- & 0.15 & -- & -- & -- & -- & 0.81 & -- \\
25 & 3.69$\times$10$^{-13}$ & 0.3 & 0.14 & 0.02 & 0.03 & 0.03 & 0.11 & 0.02 & 0.01 & 1.34 & 0.39 & 1.13 & 0.33 \\
26 & 5.74$\times$10$^{-13}$ & 0.28 & 0.13 & 0.02 & 0.03 & 0.03 & 0.1 & 0.02 & 0.01 & 1.28 & 0.37 & 1.09 & 0.33 \\
27 & 3.23$\times$10$^{-13}$ & 0.27 & 0.13 & 0.02 & 0.03 & -- & 0.09 & 0.02 & -- & 1.17 & 0.36 & 1.04 & 0.31 \\
28 & 1.28$\times$10$^{-11}$ & -- & -- & -- & -- & -- & -- & -- & -- & -- & -- & -- & -- \\
29 & 2.51$\times$10$^{-11}$ & -- & -- & -- & -- & -- & -- & -- & -- & -- & -- & -- & -- \\
30 & 3.79$\times$10$^{-11}$ & -- & -- & -- & -- & -- & -- & -- & -- & -- & -- & -- & -- \\

\enddata

\tablenotetext{a}{ erg s$^{-1}$ cm$^{-2}$. }
\tablenotetext{b}{ Fluxes are estimated to have $\pm$30\% accuracy (Section~\ref{sec:ObservationsReductions}). Propagating this uncertainty estimates the flux ratios to have $\pm$42\% accuracy. }
\tablenotetext{c}{ All other transition fluxes normalized to 1-0 S(1) value.}
\end{deluxetable}
\end{turnpage}

\clearpage

\global\pdfpageattr\expandafter{\the\pdfpageattr/Rotate 90}
\global\pdfpageattr\expandafter{\the\pdfpageattr/Rotate 0}

\begin{turnpage}
\begin{deluxetable}{lccccccccccccc}
\tablecolumns{14}
\tablewidth{0pt}
\tabletypesize{\scriptsize}
\tablecaption{Dereddened \HH~fluxes for features labeled in Figure~\ref{fig:regions} \label{table:H2_fluxes_dered}}
\tablehead{\colhead{No.} & 
                  \colhead{1-0 S(1)?} & 
                  \colhead{1-0 S(0)\tablenotemark{b}} & 
                  \colhead{1-0 S(7)} & 
                  \colhead{1-0 S(8)} & 
                  \colhead{1-0 S(9)} & 
                  \colhead{2-1 S(0)} & 
                  \colhead{2-1 S(1)} & 
                  \colhead{3-2 S(3)} & 
                  \colhead{3-2 S(5)} & 
                  \colhead{1-0 Q(1)} & 
                  \colhead{1-0 Q(2)} & 
                  \colhead{1-0 Q(3)\tablenotemark{c}} & 
                  \colhead{1-0 Q(4)} 
                  }
\startdata

1 & 2.42$\times$10$^{-14}$ & 0.24 & 0.32 & -- & -- & -- & 0.08 & -- & -- & 1.12 & -- & 0.74 & -- \\
2 & 6.88$\times$10$^{-13}$ & -- & 0.97 & -- & 0.54 & -- & -- & -- & -- & -- & -- & 0.74 & -- \\
3 & 7.42$\times$10$^{-13}$ & -- & -- & -- & -- & -- & -- & -- & -- & 0.99 & -- & 0.74 & -- \\
4 & 7.50$\times$10$^{-13}$ & 0.22 & 0.86 & -- & -- & -- & 0.12 & -- & -- & 0.94 & -- & 0.74 & 0.25 \\
5 & 1.17$\times$10$^{-12}$ & -- & 0.71 & -- & -- & -- & -- & -- & -- & 1.01 & -- & 0.74 & -- \\
6 & 1.90$\times$10$^{-13}$ & 0.22 & 0.65 & -- & -- & -- & 0.1 & -- & -- & 0.9 & -- & 0.74 & -- \\
7 & 3.29$\times$10$^{-13}$ & 0.21 & -- & -- & -- & -- & -- & -- & -- & 0.95 & -- & 0.74 & 0.22 \\
8 & 4.69$\times$10$^{-13}$ & -- & 0.54 & -- & -- & -- & 0.11 & -- & -- & 0.99 & -- & 0.74 & -- \\
9 & 6.87$\times$10$^{-13}$ & -- & -- & -- & -- & -- & -- & -- & -- & -- & -- & 0.74 & -- \\
10 & 4.45$\times$10$^{-13}$ & 0.23 & 0.33 & -- & 0.09 & -- & 0.08 & -- & -- & 0.92 & 0.26 & 0.74 & 0.15 \\
11 & 3.09$\times$10$^{-12}$ & 0.23 & 0.3 & 0.05 & 0.08 & 0.02 & 0.09 & 0.02 & 0.01 & 0.85 & 0.25 & 0.74 & 0.21 \\
12 & 1.66$\times$10$^{-12}$ & 0.23 & 0.29 & -- & -- & 0.02 & 0.08 & 0.02 & -- & 0.9 & 0.24 & 0.74 & 0.18 \\
13 & 4.08$\times$10$^{-13}$ & 0.23 & -- & -- & -- & -- & 0.1 & -- & -- & 1.11 & -- & 0.74 & -- \\
14 & 6.55$\times$10$^{-13}$ & 0.24 & 0.3 & -- & -- & -- & 0.08 & -- & -- & 0.97 & -- & 0.74 & 0.12 \\
15 & -- & -- & -- & -- & -- & -- & -- & -- & -- & -- & -- & -- & -- \\
16 & 1.38$\times$10$^{-12}$ & 0.24 & 0.3 & -- & 0.06 & 0.02 & 0.07 & 0.02 & -- & 0.85 & 0.24 & 0.74 & 0.21 \\
17 & 1.98$\times$10$^{-12}$ & 0.23 & 0.3 & -- & 0.06 & -- & 0.07 & 0.01 & -- & 0.89 & 0.24 & 0.74 & 0.19 \\
18 & 1.09$\times$10$^{-12}$ & 0.23 & 0.31 & -- & 0.08 & 0.02 & 0.07 & 0.01 & -- & 0.81 & 0.24 & 0.74 & 0.19 \\
19 & 1.37$\times$10$^{-11}$ & 0.32 & 2.06 & -- & -- & -- & 0.11 & -- & -- & 0.9 & -- & 0.74 & 0.18 \\
20 & 3.63$\times$10$^{-13}$ & 0.25 & 0.25 & -- & -- & -- & 0.1 & -- & -- & 0.84 & -- & 0.74 & 0.21 \\
21 & 7.41$\times$10$^{-13}$ & 0.26 & 0.17 & -- & -- & -- & 0.08 & -- & -- & 0.96 & 0.33 & 0.74 & 0.23 \\
22 & -- & -- & -- & -- & -- & -- & -- & -- & -- & -- & -- & -- & -- \\
23 & 2.27$\times$10$^{-14}$ & 0.29 & 0.21 & -- & -- & -- & 0.13 & -- & -- & -- & -- & 0.74 & -- \\
24 & 6.33$\times$10$^{-14}$ & 0.28 & 0.28 & -- & -- & -- & 0.15 & -- & -- & -- & -- & 0.74 & -- \\
25 & 7.69$\times$10$^{-12}$ & 0.35 & 2.07 & 0.35 & 0.67 & 0.02 & 0.12 & 0.03 & 0.03 & 0.9 & 0.26 & 0.74 & 0.2 \\
26 & 5.10$\times$10$^{-12}$ & 0.29 & 1.08 & 0.18 & 0.34 & 0.02 & 0.1 & 0.02 & 0.02 & 0.89 & 0.25 & 0.74 & 0.22 \\
27 & 1.64$\times$10$^{-12}$ & 0.24 & 0.25 & 0.04 & 0.06 & -- & 0.08 & 0.02 & -- & 0.85 & 0.26 & 0.74 & 0.22 \\
28 & 1.65$\times$10$^{-10}$ & -- & -- & -- & -- & -- & 0.09 & -- & -- & -- & -- & 0.74 & -- \\
29 & 3.72$\times$10$^{-10}$ & -- & -- & -- & -- & -- & 0.3 & -- & -- & -- & -- & 0.74 & -- \\
30 & 5.37$\times$10$^{-10}$ & -- & -- & -- & -- & -- & 0.23 & -- & -- & -- & -- & 0.74 & -- \\
\enddata

\tablenotetext{a}{ erg s$^{-1}$ cm$^{-2}$ }
\tablenotetext{b}{ All other transition fluxes normalized to 1-0 S(1) value.}
\tablenotetext{c}{ Dereddening was performed using the 1-0 Q(3)/1-0 S(1) = 0.74 intrinsic flux ratio (Section~\ref{sec:AV}; \citealt{Geballe1982}).}
\end{deluxetable}
\end{turnpage}

\clearpage

\global\pdfpageattr\expandafter{\the\pdfpageattr/Rotate 90}
\global\pdfpageattr\expandafter{\the\pdfpageattr/Rotate 0}

\begin{deluxetable}{lllr|cccccc|ccccr}
\tablecolumns{15}
\tablewidth{0pt}
\tabletypesize{\scriptsize}
\tablecaption{Observed and Dereddened \FeII~fluxes$^{a}$ and intensity-weighted radial velocities for features labeled in Figure~\ref{fig:regions}  \label{table:FeII_fluxes}}
\tablehead{\colhead{} &
                  \colhead{} &
                  \colhead{} &
                  \colhead{} &
                  \colhead{} &
                  \colhead{} &
                  \multicolumn{3}{c}{Observed} &
                  \colhead{} &
                   \colhead{} &
                  \multicolumn{3}{c}{Dereddened} &
                  \colhead{}
                  \\
                  \colhead{No.} & 
                  \colhead{RA} &
                  \colhead{Dec} &
                  \colhead{$v_{\rm LSR}^b$}  &
                  \colhead{1.26} & 
                  \colhead{1.29$^{c}$} &
                  \colhead{1.59} &
                  \colhead{1.64} &
                  \colhead{1.66} &
                  \colhead{$A_{\rm V}$} \vline & 
                  \colhead{1.26} &
                  \colhead{1.29} &
                  \colhead{1.59} &
                  \colhead{1.64} & 
                  \colhead{1.66}
                   \\
                  \colhead{} & 
                  \colhead{(J2000)} &
                  \colhead{(J2000)} &
                  \colhead{\kms} & 
                  \colhead{\mum} &
                  \colhead{\mum} &
                  \colhead{\mum} & 
                  \colhead{\mum} &
                  \colhead{\mum} &
                  \colhead{mag} \vline &
                  \colhead{\mum} &
                  \colhead{\mum} &
                  \colhead{\mum} &
                  \colhead{\mum} &
                  \colhead{\mum}
                  }
\startdata
1 & 5:35:13.704 & -5:21:49.0 & +30 & 2.42$\times$10$^{-14}$ & 0.27 & 0.17 & 1.17 & 0.12 & 14 & 1.05$\times$10$^{-13}$ & 0.25 & 0.1 & 0.67 & 0.07 \\
2 & 5:35:19.473 & -5:23:13.31 & +29 & 1.78$\times$10$^{-14}$ & 0.54 & 0.08 & 0.95 & 0.1 & 9 & 4.41$\times$10$^{-14}$ & 0.51 & 0.06 & 0.67 & 0.07 \\
3 & 5:35:11.619 & -5:20:23.66 & +7 & 1.83$\times$10$^{-14}$ & 0.4 & 0.1 & -- & 0.08 & -- & -- & -- & -- & -- & -- \\
4 & 5:35:11.625 & -5:21:11.18 & 0 & 2.46$\times$10$^{-14}$ & 0.33 & 0.14 & 1.12 & 0.1 & 13 & 9.37$\times$10$^{-14}$ & 0.3 & 0.09 & 0.67 & 0.06 \\
5 & 5:35:13.653 & -5:22:23.8 & +9 & 2.67$\times$10$^{-14}$ & 0.31 & 0.14 & 1.33 & 0.14 & 17 & 1.61$\times$10$^{-13}$ & 0.28 & 0.07 & 0.67 & 0.07 \\
6 & 5:35:12.212 & -5:20:39.56 & +12 & 2.44$\times$10$^{-14}$ & 0.31 & 0.13 & -- & 0.15 & -- & -- & -- & -- & -- & -- \\
7 & 5:35:12.96 & -5:20:40.12 & +18 & 2.79$\times$10$^{-14}$ & 0.28 & 0.08 & 1.17 & 0.09 & 14 & 1.20$\times$10$^{-13}$ & 0.26 & 0.05 & 0.67 & 0.05 \\
8 & 5:35:13.251 & -5:20:54.91 & +8 & 1.70$\times$10$^{-14}$ & 0.42 & 0.11 & -- & 0.07 & -- & -- & -- & -- & -- & -- \\
9 & 5:35:12.466 & -5:20:51.96 & +3 & 2.39$\times$10$^{-14}$ & 0.29 & 0.14 & 1.03 & 0.08 & 11 & 7.32$\times$10$^{-14}$ & 0.28 & 0.09 & 0.67 & 0.05 \\
10 & 5:35:14.092 & -5:20:38.68 & +13 & 1.05$\times$10$^{-13}$ & 0.25 & 0.07 & 0.82 & 0.06 & 5 & 1.80$\times$10$^{-13}$ & 0.24 & 0.06 & 0.67 & 0.05 \\
11 & 5:35:11.523 & -5:21:55.69 & +16 & 1.32$\times$10$^{-14}$ & 0.39 & 0.08 & -- & 0.07 & -- & -- & -- & -- & -- & -- \\
12 & 5:35:11.619 & -5:22:54.72 & +4 & 5.02$\times$10$^{-14}$ & 0.29 & 0.13 & -- & 0.08 & -- & -- & -- & -- & -- & -- \\
13 & 5:35:11.739 & -5:22:49.15 & +27 & 1.84$\times$10$^{-14}$ & 0.22 & 0.2 & 1.13 & 0.16 & 13 & 7.29$\times$10$^{-14}$ & 0.21 & 0.12 & 0.67 & 0.09 \\

\enddata
\tablenotetext{a}{ erg s$^{-1}$ cm$^{-2}$. }
\tablenotetext{b}{ Radial velocities measured only from the 1.26 \mum~transition.}
\tablenotetext{c}{ All other transitions normalized to 1.26 \mum~value.}
\end{deluxetable}


\end{document}